\documentclass[acmsmall,authorversion]{acmart}

\usepackage{enumitem}
\usepackage{subfigure}
\usepackage{graphics}
\usepackage{multirow}

\AtBeginDocument{%
  \providecommand\BibTeX{{%
    \normalfont B\kern-0.5em{\scshape i\kern-0.25em b}\kern-0.8em\TeX}}}

\setcopyright{acmlicensed}
\acmJournal{PACMHCI}
\acmYear{2021} \acmVolume{5} \acmNumber{CSCW2} \acmArticle{394} \acmMonth{10} \acmPrice{15.00}\acmDOI{10.1145/3479538}

\begin{document}

\title{Curiosity Notebook: The Design of a Research Platform for Learning by Teaching}


 \author{Ken Jen Lee}
 \authornotemark[1]
 \email{kenjen.lee@uwaterloo.ca}
 \affiliation{%
   \institution{University of Waterloo}
   \city{Waterloo}
   \state{Ontario}
   \country{Canada}
 }

 \author{Apoorva Chauhan}
 \email{apoorvathechauhan@gmail.com}
 \affiliation{%
   \institution{Concordia University of Edmonton}
   \city{Edmonton}
   \state{Alberta}
   \country{Canada}
 }
 
 \authornote{Both authors contributed equally to this research.}
 
 \author{Joslin Goh}
 \email{jtcgoh@uwaterloo.ca}
 \affiliation{%
   \institution{University of Waterloo}
   \city{Waterloo}
   \state{Ontario}
   \country{Canada}
 }
 
 \author{Elizabeth Nilsen}
 \email{enilsen@uwaterloo.ca}
 \affiliation{%
   \institution{University of Waterloo}
   \city{Waterloo}
   \state{Ontario}
   \country{Canada}
 }
 
 \author{Edith Law}
 \email{edith.law@uwaterloo.ca}
 \affiliation{%
  \institution{University of Waterloo}
  \city{Waterloo}
  \state{Ontario}
  \country{Canada}
}

\renewcommand{\shortauthors}{Lee, Chauhan, et al.}

\begin{abstract}
While learning by teaching is a popular pedagogical technique, it is a learning phenomenon that is difficult to study due to variability in the tutor-tutee pairings and learning environments.  In this paper, we introduce the Curiosity Notebook, a web-based research infrastructure for studying learning by teaching via the use of a teachable agent.  We describe and provide rationale for the set of features that are essential for such a research infrastructure, outline how these features have evolved over two design iterations of the Curiosity Notebook and through two studies---a 4-week field study with 12 elementary school students interacting with a NAO robot and an hour-long online observational study with 41 university students interacting with an agent---demonstrate the utility of our platform for making observations of learning-by-teaching phenomena in diverse learning environments.  Based on these findings, we conclude the paper by reflecting on our design evolution and envisioning future iterations of the Curiosity Notebook.  
\end{abstract}

\ccsdesc[500]{Human-centered computing~Empirical studies in interaction design}

\keywords{teachable agents; conversational interfaces; learning by teaching; research infrastructure; educational technology}

\maketitle

\section{Introduction}

Learning by teaching is a popular and well-studied pedagogical technique, shown to produce the protégé effect~\cite{Chase2009}---students learn more effectively by teaching others. When teaching, students synthesize and structure materials, become aware of their own learning process, and expend more effort to learn. Learning by teaching has been shown to benefit the tutor academically and psychologically \cite{robinson2005peer}. These benefits include improving  inquiry skills \cite{aslan2017learning}, transferring skills towards related domains \cite{robinson2005peer}, improving attitudes towards the material \cite{cohen1982educational} and school \cite{robinson2005peer}, and bolstering self-confidence \cite{srivastava2018edge}. Further, certain activities during learning by teaching have been shown to enhance such benefits, including providing explanations, collaboratively building knowledge with the tutee, and receiving feedback about the tutee's learning \cite{Duran2016,Okita2013,Okita2013_2}.  Learning by teaching has found its use in a wide variety of learning settings, at different education levels (e.g., university \cite{grzega2008didactic}, grade \cite{leelawong2008designing,chin2013young}, middle and high school \cite{Roscoe2007}) and involving different content areas (e.g., linguistics, applied sciences \cite{grzega2008didactic}, mathematics \cite{Matsuda2013} and sciences \cite{leelawong2008designing,chin2010preparing}). In education research, learning by teaching is closely related or synonymous to other terms, such as peer tutoring~\cite{robinson2005peer,Duran2016,matsuda2020effect}, cooperative learning~\cite{Duran2016}, and peer-assisted learning~\cite{garkal2018learning}.

Despite extensive research, our understanding of the specific conditions that make learning by teaching effective is limited because there are often substantial confounds introduced by the variability in the tutor-tutee pairings. One recommendation, put forth by Roscoe and Chi in a large survey on tutor learning ~\cite{Roscoe2007}, is to ``develop teachable agents to test hypotheses about specific tutor behaviours'' by systematically manipulating the teachable agent’s characteristics and question asking behaviour.  The use of teachable agents as tutees allow researchers to overcome some challenges faced in conducting studies with human tutors and tutees, including the ability to precisely control the characteristics of the tutee so that participants are exposed to the same experimental stimuli and removing the potential risk of human tutees being disadvantaged in the process \cite{Matsuda2013,matsuda2020effect,Roscoe2007}.

While many agent-mediated learning-by-teaching environments have been developed by researchers, most are not designed with the degree of flexibility required for reuse across a wide range of learning contexts  \cite{Matsuda2013,ogan2012oh,Okita2013,chin2010preparing}.  Our work seeks to fill this gap by introducing a learning-by-teaching platform called the Curiosity Notebook. Our premise is that such a research platform must be highly configurable to enable the investigation of learning by teaching in a large number of settings. This is important because, as Roscoe and Chi pointed out, tutor learning is not uniform across various settings and the reasons behind the variability are not well understood \cite{Roscoe2007}. They also encouraged future researchers to analyze and compare tutor learning processes across settings \cite{Roscoe2007}. Having a configurable research platform that supports a variety of settings and agent configurations is an important step towards facilitating comparisons across studies, by minimizing confounding factors and interactions \cite{Roscoe2007}. However, such configurability is often not found because most platforms are built to investigate learning-by-teaching phenomena in very specific learning contexts. Moreover, designing a highly configurable platform requires a significant amount of resources and is often deemed unnecessary for these narrowly focused studies. Closest to our work is Betty's Brain, where students teach a virtual agent (i.e., an avatar called Betty) about causal relationships through concept maps.  Betty's Brain has been used to study hierarchical reasoning \cite{chin2013young}, self-regulated learning \cite{biswas2005learning,Munshi2018}, collaborative teaching \cite{Emara2018}, the role of feedback \cite{leelawong2002effects,Tan2006,Segedy2012}, and scaffolded learning of multiple representations \cite{Basu2016}, to name a few;  this variety suggests a certain level of configurability allowed by the platform for researchers to adjust according to each study's specific needs.  Our work goes beyond prior work by providing configurability for agent's embodiment (e.g., text-based chatbots, voice agents or physical robots), coordinated group-based teaching (e.g., explicitly implemented turn-taking mechanisms that enable students to teach in groups of varying sizes), and learning task and materials (e.g., grade school vs. university students), all of which are discussed in this paper.

In this paper, we outline the set of configurable features in Section \ref{sec:essential_features}, informed by prior literature, that are useful for studying learning by teaching and essential for allowing better cross-study comparisons to be made.  We provide a detailed description of the platform's design evolution in Section \ref{sec:cn_design}, as informed by observations of two deployments---elementary school children participating in groups to teach physical robots in person, and university students working individually to teach a chatbot online.  These deployments serve two purposes: first, they inform potential design changes that could be made for subsequent iterations of the platform; second, they serve to demonstrate the platform's versatility as a research infrastructure in collecting useful observations for understanding learning-by-teaching phenomena in a wide range of learning contexts. In Section \ref{sec:utility}, we describe a number of observations collected using Curiosity Notebook in these deployments, as evidence for the utility of the platform and the importance of the configurability allowed by it, and conclude in Section \ref{sec:discussion} with a discussion of the envisioned design improvements for future iterations of the platform. 

\section{Background}

\subsection{Learning by teaching}
As noted above, learning by teaching produces the {\it prot\'eg\'e effect} \cite{Chase2009}---students who are asked to teach others demonstrate more proficiency in synthesizing and structuring materials, become aware of their own learning process, and expend additional effort within the learning process. The benefit of learning by teaching comes from the additional knowledge-building activities that are inherent within the process of providing explanations for others, which in turn promotes long-term understanding and consolidation of information \cite{Fiorella2015,Hoogerheide2014,Fiorella2013,Fiorella2014}. Learning by teaching also allows students to better assess their own understanding of concepts and identify the gaps in their knowledge, thereby enhancing self-learning by repairing these gaps \cite{Lachner2020}.  For instance, prior work has shown that students exhibit greater gains in knowledge when actively teaching, compared to just preparing to teach \cite{Hoogerheide2014}. 

The benefits from learning by teaching also come from learners' motivations. Benware et al.~\cite{Benware1984} hypothesized that individuals who learn material with the purpose of teaching take a more active learning role and have higher levels of intrinsic motivation. An intrinsically motivated learner engages with learning due to personal enjoyment and interests, instead of external rewards and incentives (e.g., higher test scores). According to the cognitive evaluation theory, intrinsic motivation is a culmination of self-perception of one's own autonomy (the feeling that actions can affect the surroundings) and competence (the feeling of having mastered a topic) \cite{Legault2017}. Benware et al., in \cite{Benware1984}, explained that learning for the purpose of teaching makes the learning process purposeful, as it increases autonomy which, in turn, enhances intrinsic motivation. Their study showed that, compared to participants who learn new material to take a test, those who learned to teach were more interested in the material and enjoyed the study more, supporting the hypothesis that learning to teach increases intrinsic motivation. 

As mentioned in the introduction, learning by teaching also has various other benefits, including psychological benefits. A meta-analysis by Robinson and Schofield~\cite{robinson2005peer} found that tutoring other students increases the tutor's sense of belonging to their school and with other students. Moreover, tutors who taught materials from one domain of mathematics performed better in other domains of mathematics as well, showing improvements in tutors' transferring of skills. Other studies found improvements in the tutors' self-concept (defined as the ``organization of qualities that the individual attributes to himself'' \cite{kinch1963formalized}) \cite{FloresDuran2013}, self-awareness and self-direction \cite{sprinthall1989promoting}. 

The degree to which programs involving learning by teaching succeed is influenced by numerous program characteristics, e.g., how much instruction the tutors received, age difference between tutor and tutee, how the tutors and tutees are chosen for the study. Moreover, many studies fail to tease apart the effects of teaching from other effects like collaborative learning (defined as when``two or more people learn or attempt to learn something together'' \cite{dillenbourg1999you}) \cite{robinson2005peer}. Developing \textit{teachable agents} as a method of systematically studying learning by teaching, as suggested by Roscoe and Chi \cite{Roscoe2007}, can help overcome some of these limitations.  In particular, with teachable agents as tutees, researchers have full control over the tutees' characteristics and how the tutor and tutee interact. This means that confounding factors and other closely related effects (e.g., collaborative learning) can be controlled or accounted for. More specifically, teachable agents can allow researchers to study students' responses to learning by teaching, including differential responses as the characteristics and behaviour of the agent are manipulated.

\subsection{Teachable Conversational Agents}
Within educational settings, conversational agents have adopted numerous roles. As experts, agents improve information acquisition; as motivators, agents increase self-efficacy; and as mentors, agents improve both learning and motivation \cite{baylor2005simulating}. Tutor agents have been successful in tasks including writing algebraic expressions \cite{heffernan2004web}, learning about urban ecosystems \cite{Griol2013}, and foreign language learning \cite{fryer2006bots}. 

Only a handful of systems position the agent as a less intelligent or knowledgeable "peer" that students teach. These agents are referred to as teachable agents. Using teachable agents to research learning by teaching has various benefits, including 1) enabling precisely determined control conditions not possible with human tutees, 2) reducing the risk of any human tutees from being harmed in the process, and, importantly, 3) facilitating the collection of detailed interaction data between the teacher and the agent \cite{Matsuda2013}.
SimStudent, for example, is a simulated learner used to study student-tutor learning in mathematics problem solving~\cite{Matsuda2013}. It was built with the primary purpose of collecting detailed interaction data when students are teaching the teachable agent and investigating the cognitive and social theoretic underpinnings of when, how and why learning by teaching is effective. Students teach SimStudent by posing algebraic problems to it, providing feedback and help while SimStudent attempts to solve the problems, and quizzing SimStudent to gauge learning. SimStudent was found to be effective for learning procedural, but not conceptual, skills. Further, it was more effective when students taught it correctly and made good use of the feedback and quizzing features. Another example is the Betty’s Brain system \cite{biswas2005learning,leelawong2002effects}, a teachable agent learning environment in which students read articles, then teach, query and quiz a virtual agent ("Betty") about causal relationships (e.g., burning fossil fuels increases CO2) in science by manipulating concept maps. It is designed to help students develop structured networks of knowledge, take responsibility and make decisions about learning, and develop reflection or meta-cognitive skills \cite{biswas2005learning}, thereby preparing them for future learning. 

Other researchers have explored the use of teachable agents that involve physical robots \cite{Tanaka2012a, Hood2015, Tanaka2015, Yadollahi2018, chandra2017affect} and showed the importance of incorporating recursive feedback within the design of effective agents (embodied or otherwise) for learning by teaching. Recursive feedback is when teachers observe their tutees use knowledge they taught. Okita et al.~\cite{Okita2013,Okita2013_2} found that participants who observed their tutees interacting with examiners learned better than participants who only performed learning by teaching. In another study, high-school students who observed the agents they taught compete with other agents learned more than high-school students who taught agents and proceed to compete against other agents themselves \cite{Okita2013}. It turns out that this effect is also apparent when the participants are playing the role of the examiner (instead of having a separate entity play the examiner) \cite{leelawong2002effects}. 

While teachable agents provide many benefits over human tutees, such work is labour-intensive, requiring meticulous design and implementation (i.e., coding). As such, having a research platform reusable across many learning-by-teaching studies would be significantly beneficial.

\section{Essential Features of a Research Platform for Learning by Teaching}\label{sec:essential_features}

Our goal is to create a research platform for investigating learning-by-teaching phenomena; though, certainly, there would be educational, non-research applications as well. As discussed before, a platform needs to be highly configurable to allow the investigation of learning by teaching across diverse settings. Having a single platform support multiple settings is crucial for enhancing understanding of the complex interactions in learning by teaching and for allowing cross-study comparisons (as encouraged by \cite{Roscoe2007}). Below, we discuss the configurable features (CF), their significance, and the type of research questions they support.

\textbf{CF\#1 Agent Characteristics.} Our platform enables systematic modulation of agent characteristics hypothesized to be relevant to learning (e.g., types of question asked, amount and accuracy of content demonstrated when tested). Tutee characteristics is a commonly studied and manipulated dimension within relevant literature, and is an integral part of learning-by-teaching experiences \cite{nichols1994issues,park2015boosting,biswas2005learning}. Using an agent allows for precise control over tutee characteristics and behaviours not possible with human tutees \cite{Matsuda2013,Roscoe2007}. For instance, modifying teachable agents to have them provide feedback to their teacher and ask authentic questions improves the student tutor's engagement with the material \cite{park2015boosting}. Other studied or discussed characteristics include the degree to which agents should take control of the conversation \cite{nichols1994issues} and the agents' self-regulation behaviour \cite{biswas2005learning}. Similar to how the variability in human tutees behaviours \cite{Roscoe2007} limits generalizability of findings, the variability in agent characteristics across studies makes it hard for meaningful comparisons to be made.
Thus, there is an important need for a research platform that allows for careful and granular modifications of tutee characteristics. Having configurable agent characteristics would also allow systematic analyses and verification of findings from human-to-human learning-by-teaching studies. An example is Roscoe's study that found that human tutee's questions were significant predictors of human tutors' knowledge building and deep understanding \cite{roscoe2014self}. Roscoe suggested future studies to investigate how the absolute and relative levels of tutor and tutee expertise affect the amount of knowledge building; something that is much easier to investigate using agents since it is possible to fix the level of expertise in an agent tutee, making this variable consistent across tutors \cite{roscoe2014self}.  By providing functionalities to  configure agent characteristics, our platform can support new experiments that investigate the effects of tutor-tutee pairings on learning by teaching. 

\textbf{CF\#2 Quantification of Teaching Strategies.} Our platform provides students with choices of teaching activities (e.g., teaching vs. testing), in order to allow for a quantitative characterization of their teaching strategies. This is essential as teaching strategies could be analyzed to better understand what and why certain teaching activities diminish or enhance benefits related to learning by teaching. For instance, Munshi et al.~\cite{Munshi2018} found that students who were high performers employed different patterns of teaching activities than low performers within the Betty's Brain platform.  High performing students read relevant articles after quizzing Betty more frequently than low performing students. Using the same platform, Wagster et al.~ \cite{wagster2007learning} analyzed how high- and low-performing students employ the six available teaching activities while teaching the agent. They found that high-performing students incorporated more activities related to information seeking and self-monitoring than low-performing students. Beyond Betty's Brain, however, there is a lack of research on the effects of teaching activities and strategies on the learning of the tutor teaching an agent tutee; as such, our understanding of these effects is limited to tasks for learning causal relationships between concepts, which Betty's Brain supports.  An important configurable feature of our platform, therefore, is a set of user interface controls (e.g., buttons) that student tutors can invoke to initiate a range of different teaching conversations with the agent. These user interface controls enable researchers to observe what teaching activities the student tutors would choose to engage in and in what order, thus providing rich data for capturing learning-by-teaching behaviour. This can help towards understanding not only \textit{what}, but \textit{how} and \textit{why} tutors learn from teaching, which have been identified as notable gaps within the extant research studies \cite{Roscoe2007}.

\textbf{CF\#3 Scalable Learning Task and Material.} Our platform supports learning tasks that can be scaled in complexity to different age groups (e.g., usable by elementary school as well as university students). Existing learning-by-teaching research involving teachable agents (i.e., agent tutees) are mostly for younger children (usually in elementary school) \cite{leelawong2008designing,Emara2018,wagster2007learning,biswas2005learning,Munshi2018,leelawong2002effects}. Research on learning by teaching for older students (e.g., university students), on the other hand, mostly investigated human tutor/tutee contexts without the use of teachable agents \cite{annis1983processes,Coleman1997,roscoe2004influence,macdonald1991analysis,roscoe2014self}.
Research have also shown that the learning process itself evolves with age. For example, although the associative learning process might remain constant in children and adults, executively controlled learning (i.e., learning that requires engagement on the learner's part) increases in frequency with older ages \cite{kuhn2006children}. Metacognitive skills also develop with age; adults are more capable of such skills than children, including taking more initiative to verbally rehearse during memory tasks \cite{keeney1967spontaneous}, improve on their retrieval skills given previous failed attempts \cite{flavell1975metamemory}, and employ knowledge acquisition strategies \cite{kuhn2000metacognitive}.  As such, while learning by teaching has been found to benefit both children and adults alike, some important questions remain unanswered: 1) are teachable agents as effective as human-to-human learning by teaching for older students? and 2) how is the effectiveness of teachable agents for young children similar or different for older students? Having a single platform that supports multiple age groups and learning task complexities allows for a direct comparison between these age groups, and eventually allowing learning by teaching to benefit students of different ages through informed and targeted configurations.

\textbf{CF\#4 Coordinated Group-Based Teaching.} Our platform supports students teaching individually, in pairs or in larger groups, explicitly controlling the turn-taking process in order to provide equal access to teaching opportunities. Collaborative learning, defined as ``the grouping and pairing of learners for the purpose of achieving a learning goal'', has many benefits \cite{laal2012benefits}, which include encouraging critical thinking and increasing students' motivation, and promoting active student participation and personalized learning \cite{laal2012benefits}. 
In teachable agents contexts, however, the only studies comparing individual versus collaborative learning by teaching, are those using Betty's Brain.  Namely, Emara et al.~\cite{Emara2018} found that when paired to teach Betty together, students learned more and created more accurate concept maps than when working individually. This effect emerged because working in pairs allowed students to discuss errors in quizzes and their concept maps, which in turn encouraged more systematic approaches to learning and teaching the agent \cite{Emara2018}. While the findings are significant, many questions remain about the inner working of learning by collaboratively teaching teachable agents. In particular, it is not well understood how a group's dynamics affect each tutor's learning. Within collaborative learning literature, researchers found that its effectiveness is affected by students' personality (e.g., extroverted vs. introverted), cognitive style \cite{miller1994group}, motivation levels \cite{rienties2009role}, and the group's contribution equity \cite{shah2014analyzing}. For instance, Shah et al.~\cite{shah2014analyzing} found that students who perceived themselves as being less competent in the subject were dominated in discussions within dyads, and as a result, were negatively affected in their learning process. These results raise questions of how teachable agent systems should be designed to manage group dynamics, e.g., encourage or enforce equal engagement and contribution among tutors.  By providing functionalities that coordinate  group-based teaching, our platform enables new questions about collaborative learning by teaching to be answered. 

\textbf{CF\#5 Flexible Agent Embodiments.} Our platform supports teaching conversations with text-based, voice-based or physical agents. Agent embodiments, beyond simple text interfaces (e.g., auditory speech, digital or physical animated forms), have been found to improve performance in tasks such as tutoring, where continuous interactions with agents are held \cite{shamekhi2018face,atkinson2002optimizing,moreno2001case}. However, agent embodiment is nuanced and rarely investigated in a thorough manner; these nuances are often listed as limitations in studies~\cite{hone2003affective,shamekhi2018face,vossen2009social,thellman2016physical}. Although out of this paper's scope, supporting flexible agent embodiment would allow for fair and detailed comparisons of the effects of embodiment on teaching and learning performance. For instance, one can investigate how agent embodiment affects rapport and how increased/decreased rapport would impact  learning-by-teaching behaviours and outcomes. 

To summarize, the five configurable features above are integral for researching learning by teaching with teachable agents. However, as discussed above, many questions remain unexplored for each configurable feature, and most studies use different learning-by-teaching platforms, making it hard to compare and synthesize generalizable findings that can be easily applied to the education sector. Although prior work has explored a few of the configurable features in depth, the effects of CF\#3 and CF\#5 remain unexplored. As such, we set out to build Curiosity Notebook to support all five configurable features, and introduce its design in the next section.

\section{Curiosity Notebook: a Research Platform for Learning by Teaching}\label{sec:cn_design} 

In this paper, we introduce the Curiosity Notebook---a new research platform for learning by teaching---and describe how its design evolved over two years of development. We also demonstrate how the platform supports each configurable feature (e.g., CF\#3) discussed in Section \ref{sec:essential_features}.

\begin{figure}[htbp!]
  \centering
  \fbox{
  \includegraphics[width=0.75\linewidth]{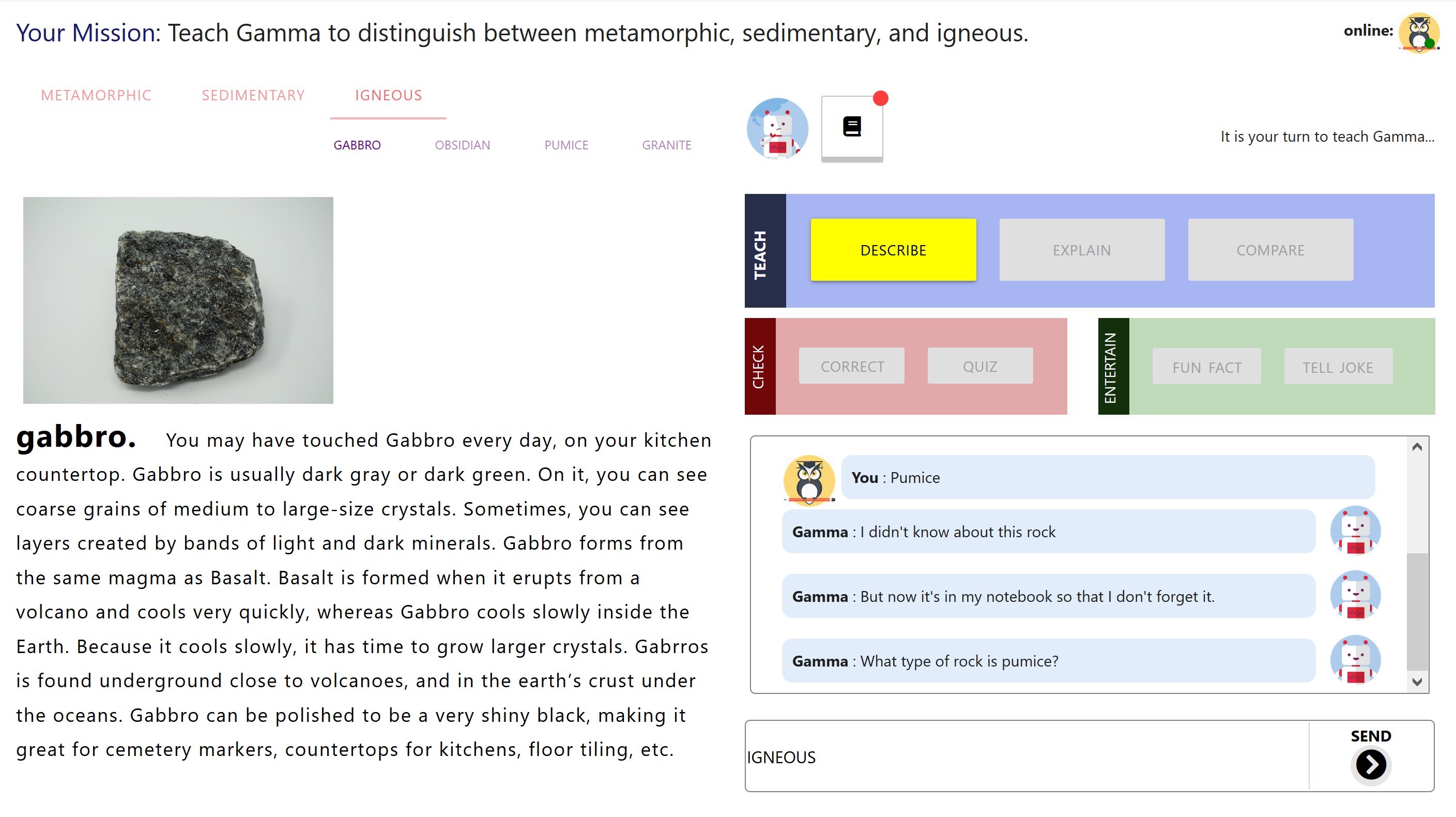}}
  \caption{Curiosity Notebook (second version) with reading panel (left) and teaching panel (right). 
  }
  \Description[Design of Curiosity Notebook]{This picture shows the Curiosity Notebook. Seven buttons representing unique conversations were added. The agent also has a notebook where it can record notes of what it has been taught.}
  \label{fig:teaching_interface}
\end{figure}

At a high-level, the Curiosity Notebook provides a web interface that students use to read articles and teach an agent to perform a task based on the reading.  In our design thus far, we have focused exclusively on how the Curiosity Notebook enable the teaching of classification tasks---how to classify objects, e.g., animals as mammals, insects, and reptiles; rocks as sedimentary, igneous and metamorphic (rocks have also been used as a topic in previous studies on agents for education \cite{Ceha:2019:7}); or paintings as pointillism, realism and impressionism. However, the task could take many forms; for example, students could read an article about programming and teach the agent how to correct mistakes in a piece of code, or students could read an article about grammar rules and teach the agent how to identify part of speech in a sentence.  The main idea is that the platform supports the process of students teaching factual knowledge (obtained from reading) in a way that builds up and transforms into a particular skill (e.g., the ability to apply rules to classify objects) in the agent.  For classification, the agent would need to be able to identify features of an object, identify the categories that the object can belong to, and map features to categories.  The platform, as described in detail below, is designed to facilitate this text-to-rule translation.  Shown in Figure \ref{fig:teaching_interface}, the main interface consists of a reading panel (left), and the teaching panel (right), which provides functionalities that enable students to communicate with the agent and assess the progress of the agent's learning.  

\subsection{Deployments}\label{sec:deployments}

During our iterative design process, we deployed two versions of the Curiosity Notebook, each to a vastly different user population and learning setting.  Before describing Curiosity Notebook's design evolution, we first describe the details about the two deployments.   

The deployments served two purposes---first, to inform the iterative design changes in the platform, and second, to demonstrate the platform's versatility as a research infrastructure. For instance, the two deployments described below demonstrate how the platform can support users from CF\#3 different age groups (elementary school children vs. university students), learning contexts (in-person vs. online), CF\#4 group configurations (group-based vs individual teaching) and with CF\#5 different agent embodiments (e.g., robot vs. chatbot). The differences between the deployments are summarized in Table \ref{table:compareDeployments}.\\

\begin{table}[htbp!]
   \caption{Comparison of deployment characteristics.}
    \label{table:compareDeployments}
  \begin{tabular}{p{0.32\textwidth}p{0.25\textwidth}p{0.25\textwidth}}\\
    \toprule
    \multicolumn{1}{c}{ } & \multicolumn{1}{c}{Deployment 1} & \multicolumn{1}{c}{Deployment 2}\\
    \midrule
    Setting & In-person within school & Online\\
    Length & 4 weeks, 1.5 hours per week & Single session, 1 hour\\
    Education level (CF\#3) & Elementary school & University \\
    & (Grades 4 \& 5) &\\
    Number of students per agent (CF\#4) & 3 & 1\\
    Agent embodiment (CF\#5) & Physical Robot (NAO) & Chatbot\\
  \bottomrule
\end{tabular}
\end{table}

\begin{figure}[ht]
  \centering
  \includegraphics[height=4cm]{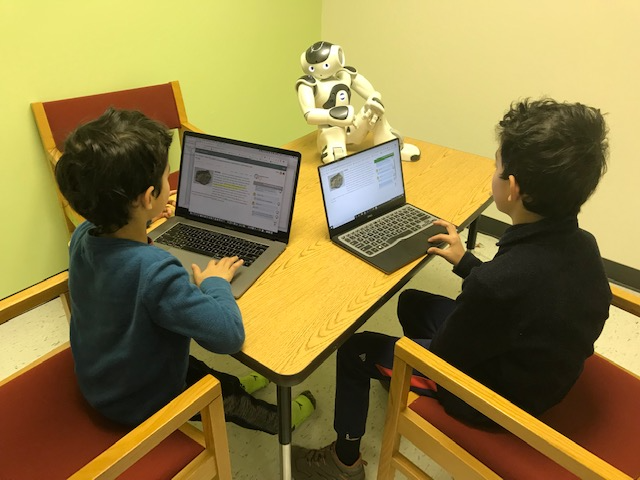}
  \includegraphics[height=4cm]{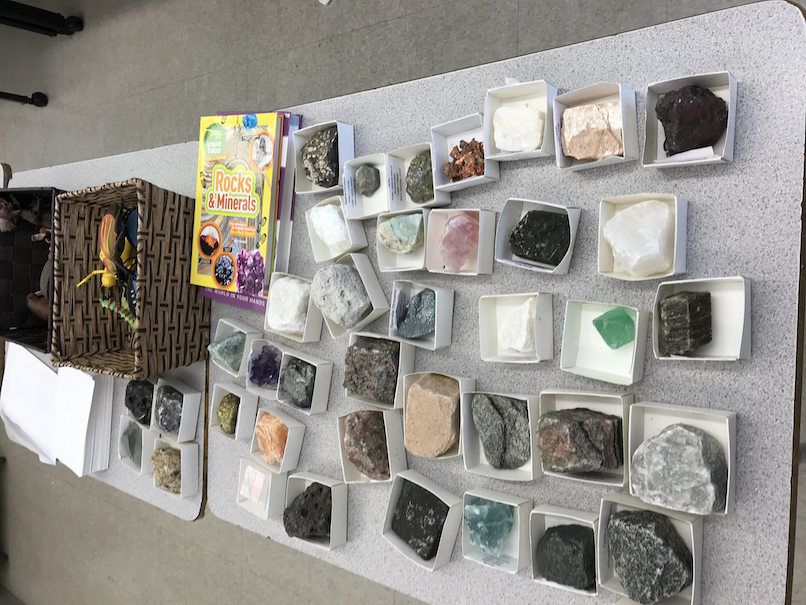}
  \caption{Students teaching a NAO robot (left) using the Curiosity Notebook and physical artifacts (right).}
  \Description[Images of field study]{There are two pictures side by side. The left picture shows two school children using the Curiosity Notebook simultaneously on separate laptops while seated next to each other. A NAO robot is placed in front of the children to communicate with them. The right picture shows artifacts of various rocks placed on a table. These rocks were used by the school children during the study.}
  \label{fig:exploratory}
\end{figure}

\noindent {\bf Deployment 1: Elementary School Students, In-Person.}  For the Curiosity Notebook's first version, we conducted a 4-week exploratory study with 12 fourth- and fifth-grade students (7M/5F) at a local school \cite{law20}. Enrollment was on a first come, first served basis. No monetary compensation was provided; instead, students were given a ``Certified Robot Teacher'' certificate as a token of appreciation. The study was conducted in an after-school club, which ran once a week for 1.5 hours each. Four NAO robots were used in each session; to personalize the experience, each robot had a name tag hung around their neck with a gender-neutral name (i.e., Alpha, Beta, Gamma and Delta). Students formed groups of three, and taught the robot about a different topic (i.e., animals, rocks, paintings) each week, then all topics during the last week. Each student was given a Chromebook, and sat together with their group members facing the robot, which was positioned in a sitting posture in front of the students on the table (as shown in Figure \ref{fig:exploratory}). Each group of students was joined by a student researcher, who observed the group and resolved issues (if any) with the platform.  We provided physical artifacts (as shown in Figure \ref{fig:exploratory}) for each classification task, namely animal figurines, rocks and minerals, and postcards of different styles of paintings from NYC Metropolitan Museum. During the session, we piloted a variety of surveys, iteratively re-designed the platform, made detailed observations, and interviewed students about their learning-by-teaching experience.\\ 

\noindent {\bf Deployment 2: University Students, Online.}  After the first deployment with elementary students, we revised the design of the Curiosity Notebook, and deployed it in a second observational study with university students.  After a small pilot with 11 participants, we recruited 41 participants (22M, 19F; ages 18 to 29, $Mdn = 21.0$).  All studies were conducted over Zoom in the presence of a researcher and were about 90-minutes long. Participants were given an \$20 CAD Amazon gift card as remuneration. Participants were asked to read articles about rocks and teach the agent (named Gamma) how to classify rocks as metamorphic, sedimentary, or igneous.  We recorded all interactions (such as article clicks, button clicks, notebook checks, text logs of chat data between participant and the agent, and the amount of time a participant spent teaching Gamma) on the Curiosity Notebook as well as participants' responses to the administered pre-study and post-study questionnaires (submitted via Google Forms). The pre-study and post-study questionnaires measured participants' attitudes towards the agent (via likeability and perceived intelligence subscales of the Godspeed questionnaire \cite{Bartneck2008}), and towards the teaching task (via Pick-a-Mood pictorial self-report scale \cite{Desmet2016} for self and agent). The Interest/Enjoyment, Pressure/Tensions and Effort/Importance subscales from the Intrinsic Motivation Inventory (IMI) \cite{ryan1982control} were used for a deeper understanding of participants' subjective experiences with the teaching task. The Academic Motivation Scale (AMS) was used to measure participants' amount and type of motivations towards the experiment \cite{vallerand1992academic}. Lastly, open-ended questions about participants' perceptions of the agent and the Curiosity Notebook were added for additional qualitative insights.

Beyond this paper, our platform has been used in a number of studies investigating the effects of agent characteristics (CF\#1) on learning by teaching---to investigate agents with different humour styles in relation to student effort in teaching \cite{cehaHumour}, and to study physically embodied agents that are capable of sensing group dynamics and verbally encouraging equal contributions among student tutors (which is also relevant to CF\#4 and CF\#5) \cite{ravarieffects}. The flexibility of the Curiosity Notebook platform (in the form of the five CFs) not only allows for its deployment in an expanded number of settings consisting of varying characteristics, but also allow researchers to expeditiously switch between these settings.

\subsection{Design Evolution}

We deployed the first version of the Curiosity Notebook to the first exploratory study with elementary school students, revised the design based on our observations and findings, then deployed the second version to the second exploratory study with university students.  There were a number of features that evolved substantially between the two versions.  We describe here these platform features and their design evolution. 

\subsubsection{Teaching Conversations}

\begin{figure}[htbp!]
  \centering
  \subfigure[First Version: knowledge bubbles indicating which features have been taught.]{%
  \label{fig:first-teachingpanel}%
  \includegraphics[height=6cm]{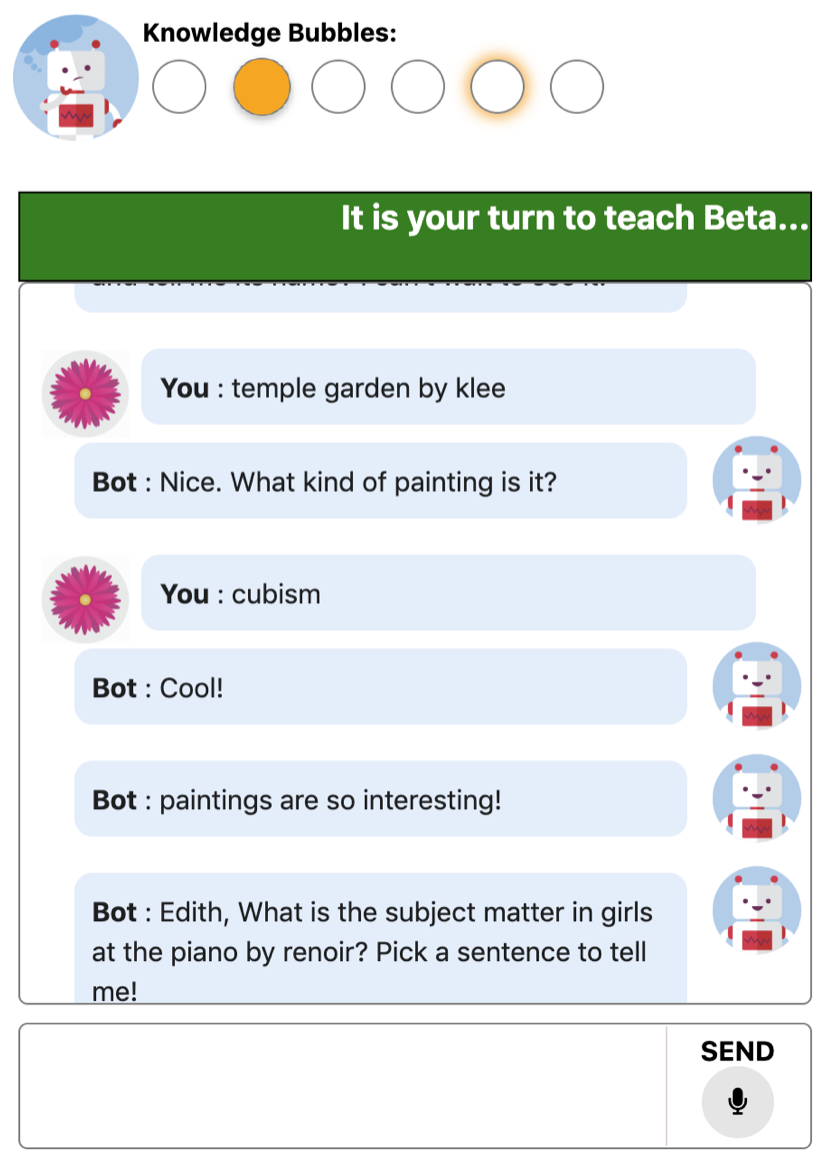}}%
  \qquad
  \subfigure[Second Version: buttons for initiating teaching conversations and robot's notebook for tracking learned knowledge]{%
  \label{fig:second-teachingpanel}%
  \includegraphics[height=6cm]{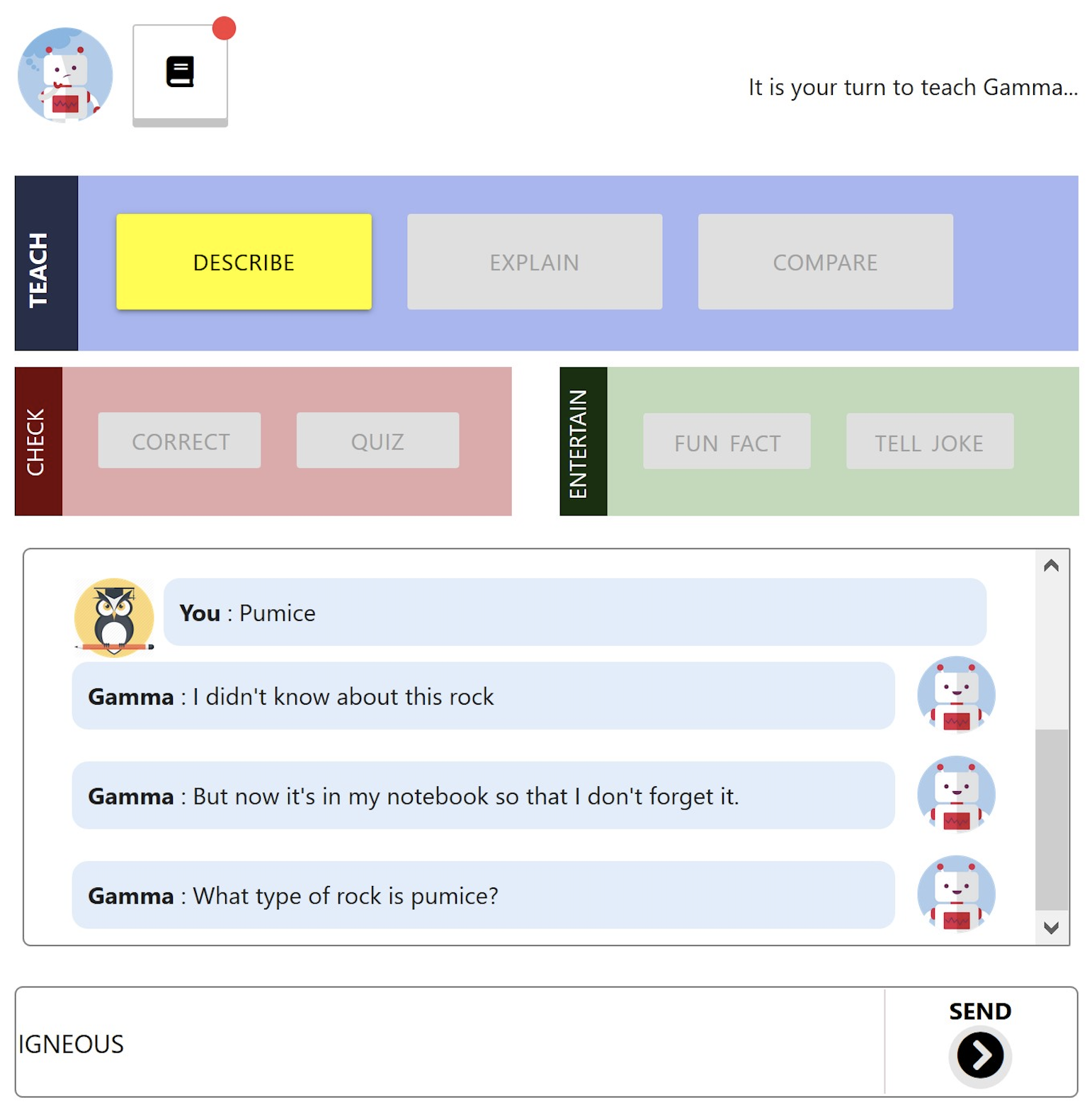}}%
  \qquad
  \caption{Design Iterations of the Teaching Panel}
  \label{fig:design-teachingpanel}
\end{figure}

In the first version of the Curiosity Notebook, the agent begins each teaching conversation by asking the students to show it a physical artifact (e.g., ``Can you pick an animal and tell me its name? I can’t wait to see it!''). After selecting an object, the teaching conversation proceeds with the agent highlighting one of the ``knowledge bubbles'' and asking a series of 4 or 5 questions about the corresponding feature, as shown in Figure \ref{fig:first-teachingpanel}. 

The agent is designed to ask a mix of low- and high-level questions. Low-level thinking questions include questions about features (e.g., ``What does the skin of mammals look like?''), examples (e.g., ``Can you give me an example of a cubism painting?''), and facts (e.g., ``Select a sentence to tell me about frogs and how they lay eggs.''). High-level thinking questions include why questions (e.g., “Why is a snake a reptile?”), synthesis questions (e.g., “Do all reptiles look the same?”), and questions that prompt students to repeat/rephrase or explain the meaning of a word, e.g., “Can you help me understand what you just said better?”  These questions reflect different categories of cognitive operations posited by Gallaghar and Aschner \cite{gallagher1963preliminary}. While low-level questions involve cognitive memory operations via the reproduction of factual knowledge directly from the articles, high-level questions involve convergent thinking via the discovery and learning of concepts (which take the form of classification rules in the context of classification tasks) from analyzing and integrating knowledge derived from the articles \cite{gallagher1963preliminary}; convergent thinking questions have been shown to be beneficial to students' question-asking skills \cite{mehdi20}.  The system automatically generates a variety of questions by filling in predefined sentence templates with names of objects, features, and categories that the students are currently teaching. The questions are sequenced such that lower-level thinking questions always precede higher-level thinking questions, allowing students to first learn (through teaching) factual knowledge before concepts relating to classification rules. Some randomness was introduced for ordering questions to prevent the conversation from being too mechanical. The teachable agent occasionally seeks feedback from students about its learning, by asking questions about its general intelligence (e.g., ``Am I smart?''), its learning progress (e.g., ``Am I learning?'', ``Do you think I know more now than before?''), or how well it might perform if tested (``Will I do well in a test?'').

After 4-5 rounds of questioning and answering, a ``knowledge bubble'' is filled and students are rewarded with confetti on the screen letting them know that the agent has ``learned'' that feature, and the next feature is randomly chosen for students to teach.  Between teaching conversations, students can also choose to test the agent’s knowledge. The testing interface shows a set of images representing objects (e.g., images of paintings) to be classified. Students can click on an image, and the agent will attempt to classify it (e.g., saying ``I think it is an impressionist painting'' or ``I don’t know''). The system provides corrective feedback—an overlay over each image will show a green checkmark if the agent is correct, and a red ``x'' otherwise. This testing feature builds on prior research, which shows that student tutors learned more when observing their tutees use the knowledge they were taught either through some form of quiz or test \cite{Okita2013,Okita2013_2,leelawong2002effects}, or through competition with other tutees \cite{Okita2013}.

A key change in the second version's design is a set of controls that allows participants to choose specific actions to take next; this, in turn, allows us to more quantitatively capture the variety of teaching strategies employed (CF\#2). In lieu of the knowledge bubbles, the teaching panel (as shown in Figure \ref{fig:second-teachingpanel}) now contains, at the top, seven ``conversation-type'' buttons grouped into three categories, supporting \textit{Teach}, \textit{Check} and \textit{Entertain} actions. When a button is clicked, the system locks the agent (named Gamma) and the participant into a conversation that involves 4-6 rounds of question and answering.  The student tutor can engage in the conversation through free-form text, or by highlighting specific sentences in the text to teach the agent.  

\begin{table}[htbp!]
\caption{\textit{Explain} conversation in the second version of the Curiosity Notebook}
\label{tab:ExplainConversation}
{\small
\begin{tabular}{l p{11cm}}
{\bf Gamma:} & It's good to understand better why rocks look the way they do.\\
{\bf Gamma:} & Can you pick a new rock and tell me what it's called please? \\ 
{\bf User:} & Shale \\ 
{\bf Gamma:} & I don't know about this rock.\\
{\bf Gamma:} & But now it's in my notebook so that I don't forget it.\\ 
{\bf Gamma:} & What category does shale belong to? \\
{\bf User:} & Sedimentary \\ 
{\bf Gamma:} & Can you tell me why shale looks the way it does?\\ 
{\bf Gamma:} & Can you pick a sentence in the articles for me please? \\ 
{\bf User:} & As more sediments get deposited, the particles underneath become tightly packed; eventually, they become a dense, solid rock. \\
{\bf Gamma:} & Mmm.. okay I see.\\ 
{\bf Gamma:} & Could you explain that more clearly? \\
{\bf User:} & With time, sediments get deposited over each other, forming a dense solid rock. \\
{\bf Gamma:} & Alright.\\ 
{\bf Gamma:} & I'm really interested in rocks.\\
{\bf Gamma:} & You can now select a new button to keep teaching me.
\end{tabular}}%
\end{table}

Each button initiates a teaching conversation with the agent. {\it Teach} actions include {\it Describe}, {\it Explain} and {\it Compare}.  Upon clicking the {\it Describe} button, the agent will ask participants to identify an object's category (e.g., ``Pumice is an igneous rock.'') and feature (e.g., ``Pumice is often white.''). This conversation allows students to better grasp factual knowledge, which is an essential step in developing competence in the topic taught \cite{national2004students}. The {\it Explain} conversation prompts for an explanation for why an object has a particular feature (e.g., ``Pumice is often white because of the silica in the lava; without the silica, Pumice could be black.''). A sample {\it Explain} teaching conversation is shown in Table \ref{tab:ExplainConversation}. Providing explanations is an important teaching activity that has proven to improve learning significantly, especially while learning the material \cite{Lachner2020}. The {\it Compare} conversation allows the participant to discuss similarities or differences between two objects. Ziegler and Stern found that when encouraged to make comparisons between conceptually different material, students benefited from larger long-term learning gains \cite{ziegler2014delayed}. In terms of cognitive operations, \textit{Describe} supports cognitive memory while \textit{Explain} and \textit{Compare} support convergent thinking \cite{gallagher1963preliminary}.  

\begin{figure}[htbp!]
  \centering
  \fbox{\includegraphics[height=8cm]{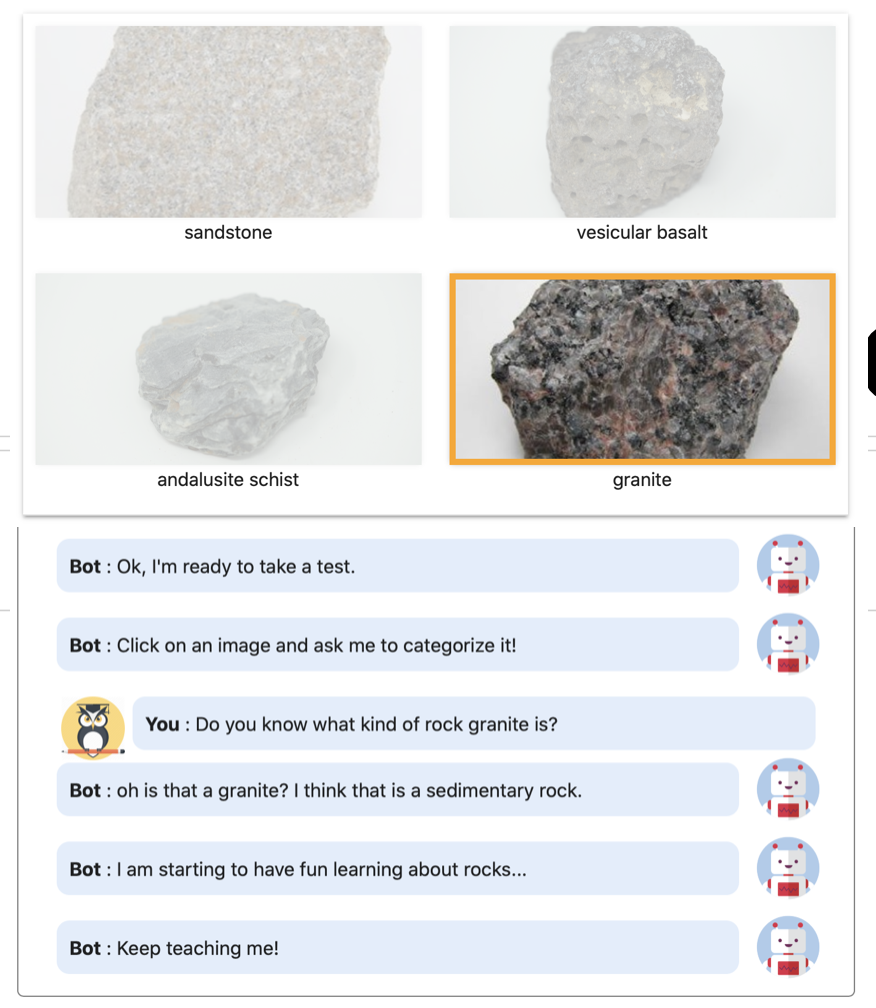}}
  \caption{\textit{Quiz} conversation in the second version of the Curiosity Notebook}
  \Description[Example of quiz conversation]{The picture shows an ongoing quiz conversation. The agent starts by saying "Okay I am ready to take a test. Click on an image and ask me to categorize it!" The user is then presented with pictures of four different rocks above the chat window. The user clicks on the picture of granite. This creates a new message in the chat window by the user, saying "Do you know what kind of rock granite is?" The agent replies by saying "Oh is that a granite? I think that is a sedimentary rock. I am starting to have fun learning about rocks. Keep teaching me!" This concludes the quiz conversation.}
  \label{fig:quiz}
\end{figure}

{\it Check} actions include {\it Correct} and {\it Quiz}. The {\it Correct} button allows the student to correct facts that they deemed the agent to have learned incorrectly (due to them teaching incorrectly; the agent in this version is not configured to inject errors automatically). This feature is designed to encourage metacognitive skills among student tutors. Specifically, by monitoring and identifying mistakes in the agent's understanding, student tutors may also identify similar flaws in their own understanding, since the agent learns directly from them. This is referred to as comprehension-monitoring, and has been found to be an essential part of knowledge building in learning by teaching \cite{roscoe2014self}. To probe the current performance of the agent, students can click on the {\it Quiz} button, and select an object to test the agent on (Figure \ref{fig:quiz}). The agent will classify the object correctly or incorrectly, based on its current knowledge model. As with the first version, the quiz feature encourages learning by providing feedback to the student tutors about the agent's learning progress.

Finally, participants can choose to entertain the agent through the {\it Entertain} actions.  The {\it Fun Fact} button allows the participant to tell a fun fact about an object (e.g., ``Obsidian is used in heart surgery'') and provide a reason for why the fact is interesting. The {\it Tell Joke} button allows the participant to tell the agent a joke. Powell and Andersen found that the use of humour in learning settings, if not excessive, can benefit learning by increasing students' attention and motivation \cite{powell1985humour}.

Together, these seven conversations are designed to support flexible teaching strategies on the side of the student tutors, and quantifiable teaching strategies that can be analyzed on the side of the researchers (CF\#2). A final key difference between the first and second version of the Curiosity Notebook is the dialog system that drives the conversation.  In the first version, the question templates were randomly drawn; whereas in the second version, each conversation is controlled by a state machine, of which configurations are stored in a JSON file that specifies the conversation's flow. This setup enables the conversations to be easily modified without having to change any part of the code (CF\#1).


\subsubsection{Coordinated Turn-Taking}

Our platform is highly configurable and supports a wide variety of learning-by-teaching scenarios.  Students can teach the agent individually or in groups of arbitrary size, and their group placement can be configured by teachers or researchers through a command line interface in the first version, and through an admin interface in the second version (CF\#4). If a student is placed in a group and their group members are present, their view of the system is synchronized—that is, if one student navigates to another interface (e.g., teaching vs. testing), all students will be automatically brought to the same screen.  During the first deployment, between each after-school session, there were design modifications to the turn-taking mechanisms.  Initially, our platform gave students complete freedom to choose what and when to teach the robot. This setup was too open-ended, and students had great difficulty narrowing down what content mattered and dividing the teaching task. Subsequently, the agent was redesigned to control turn taking---namely, it determines which group member is online and active, asks the student who has participated the least number of turns to teach next. When a student is stuck (e.g., picked a sentence unrelated to what the agent is asking about), the agent will also delegate the task to the next student, asking them to help. Overall, the turn taking mechanism enabled children, placed in different-sized groups within the same classroom, to work together on simultaneously teaching different robots.   


\subsubsection{Agent Embodiment Logic}

The Curiosity Notebook supports a clean separation between agent logic and embodiment, thereby allowing the teachable agent to take on different types of embodiment (CF\#5). This is accomplished by keeping the teachable agent's logic---e.g., how it learns, how it exhibits emotions, and what it says—inside the Curiosity Notebook web application, and having an external program (e.g., a python script) ping the database for chat messages that the physical robot should say out loud. Each chat message is associated with an emotion tag (e.g., curious), which can be used to control the movements/gestures of the robot (e.g., rubbing its head or chin) to convey that emotion. Similarly, the external program can push sensing events to the Curiosity Notebook. The NAO robot, for example, has tactile sensors on its head, hands and feet, which can serve as alternative ways for students to provide feedback to the robot (e.g., patting its head when it answers a quiz question correctly). If the Curiosity Notebook is used without any physical embodiment for the agent, the agent takes the form of a text-based conversational agent.  This functionality enabled us to set up the agent to be a physical robot in the first deployment, and a chatbot (i.e., text-based conversational agent) in the second deployment. In other words, supporting flexible agent embodiments allows for studies that use different embodiments to be deployed either simultaneously or in quick succession without having to change the platform's core logic.


\subsubsection{Interfaces for Monitoring the Agent's Knowledge} \label{sec:notebook}

As shown earlier in Figure \ref{fig:design-teachingpanel}, in the first version of the Curiosity Notebook, the state of the agent’s learning is represented by knowledge bubbles, each representing a feature that is relevant for the classification task at hand. For example, a feature relevant for distinguishing mammals and reptiles would be whether the animal lays eggs. A knowledge bubble becomes filled when an agent has mastered/learned the associated feature, otherwise it remains empty. In our first deployment, we found that although filled knowledge bubbles were reflective of the agent's knowledge, students could not gauge from the bubbles what content the agent learned and how different actions affected the agent's knowledge. The knowledge bubbles also had the unintentional effect of encouraging students to rush through teaching in order to fill up all the knowledge bubbles as quickly as possible, and reducing their focus on learning the material. 

\begin{figure}[htbp!]
  \centering
  \fbox{\includegraphics[height=6cm]{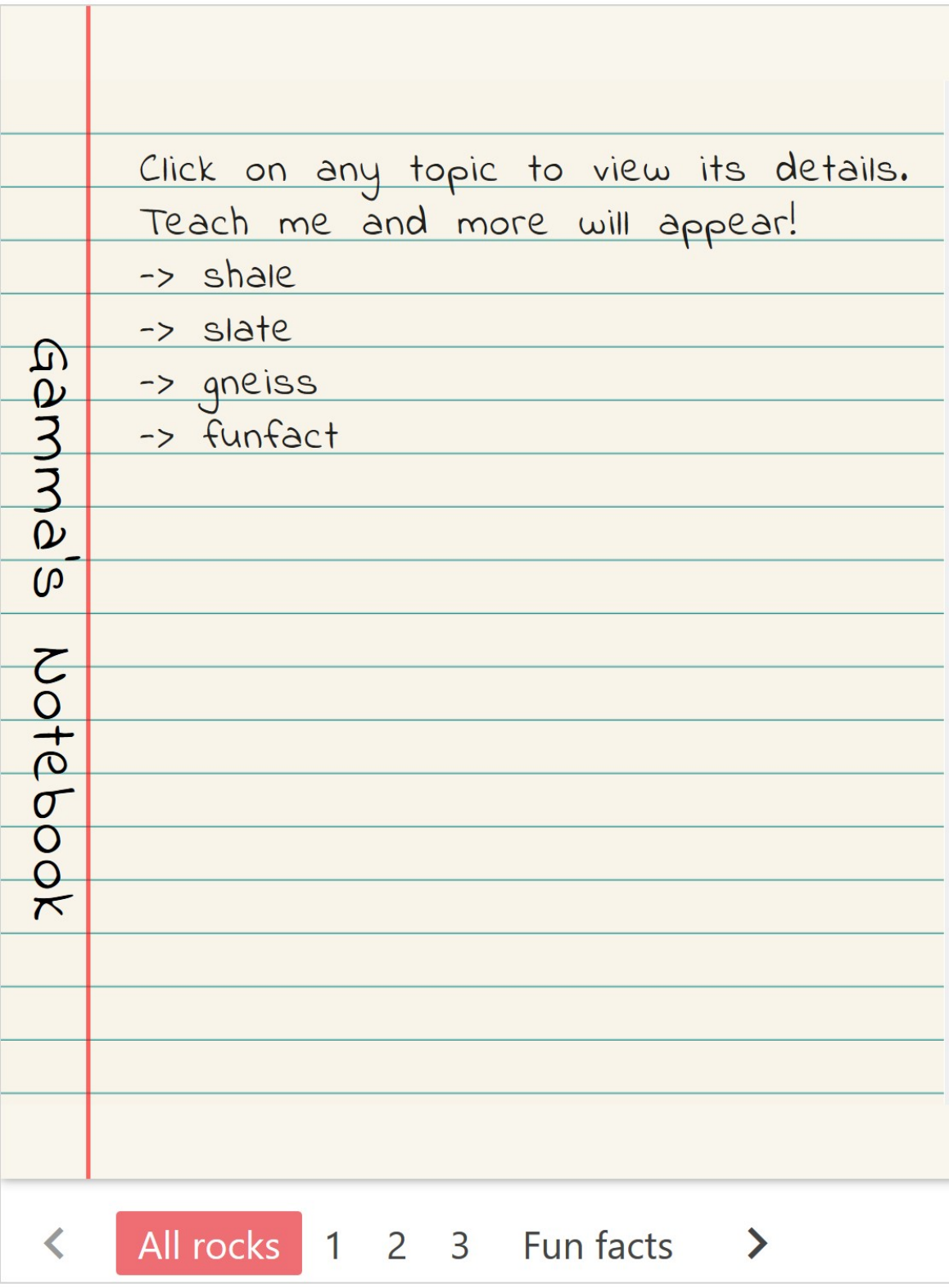}}
  \hspace{1cm}
  \fbox{\includegraphics[height=6cm]{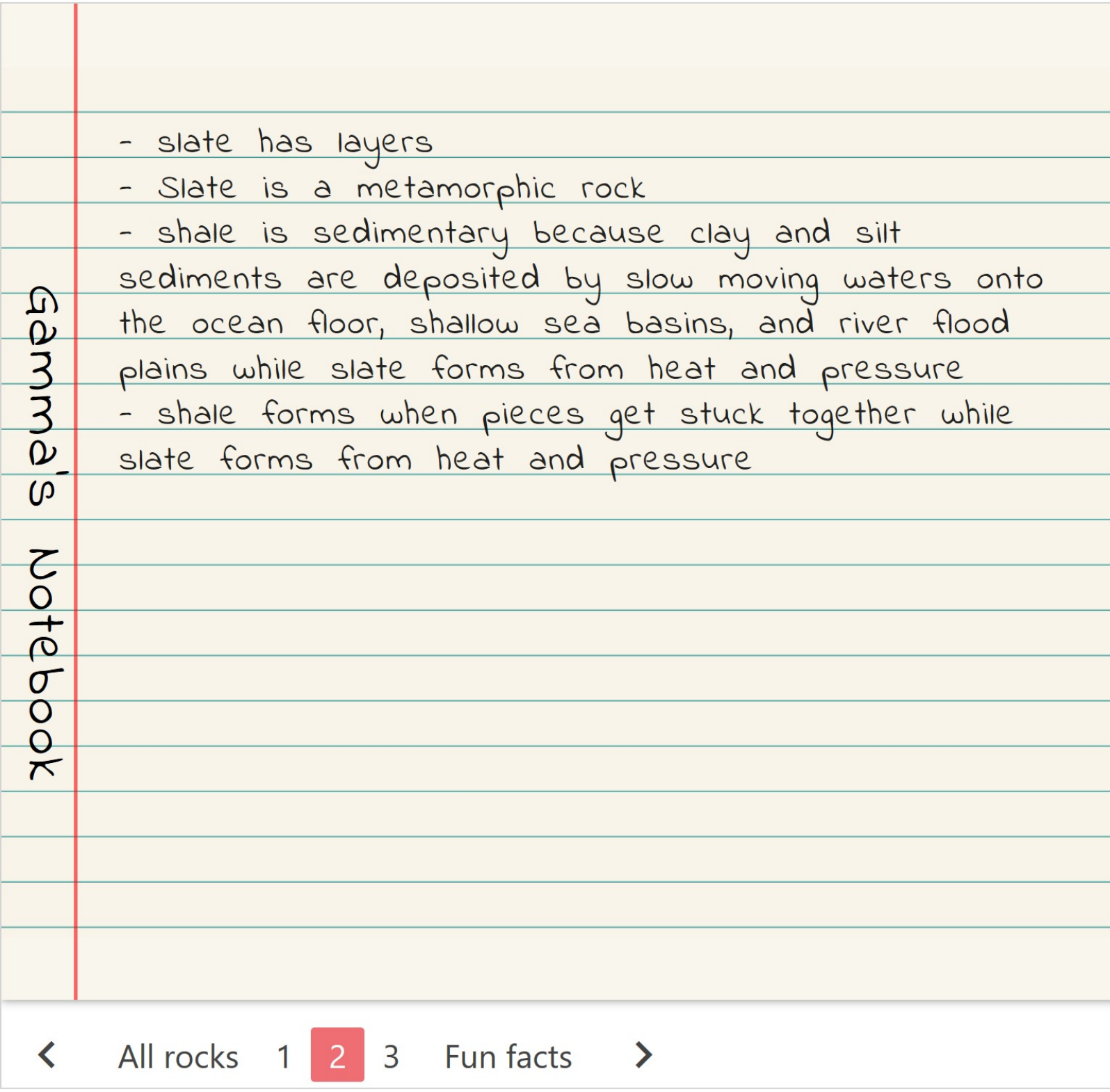}}
  \caption{Table of content page of the agent's notebook (left) and the page containing notes for Slate (right), in the second version of the Curiosity Notebook}
  \Description[Images of notebook interface]{There are two images showing the agent's notebook. The left one shows the index page of the notebook, which has names of the rocks that the agent has been taught. The right picture shows a page for the rock Slate containing Slate related notes.}
  \label{fig:notebook_interface}
\end{figure}

To address these issues, in the second version, the knowledge bubbles were replaced with the agent's notebook (Figure \ref{fig:notebook_interface}), which records everything that the agent has been taught. The agent's notebook is informative in two ways; it tells students what the agent knows, and it gives students clues as to what each button does.  For example, sharing a joke with the agent does not affect its knowledge, but telling a fun fact does. Throughout teaching, students can access the agent's notebook using the \textit{Notebook} button (next to the agent's avatar in Figure \ref{fig:second-teachingpanel}) at any time. In the initial testing, we found that users who were not familiar with the interface often forgot that the agent's notebook existed. Thus, extra logic was built to make the \textit{Notebook} button pulsate for approximately two seconds every time a new note has been taken by the agent. Additionally, the agent is programmed to tell the user through chat that it has made a new entry in its notebook.

The notebook starts out empty initially, with the first page serving as a table of content. Upon learning a fact about a rock, a new page is created for that rock with the added note, and a new index to the page is added to the table of content. The last page of the notebook is reserved for showing all the fun facts.  The cursive Indie Flower font~\footnote{https://fonts.google.com/specimen/Indie+Flower} was chosen for the notebook texts to represent the agent's handwriting.  The information that the agent learns is stored in a relational database.  For instance, for rock classification, the rock-to-category table stores what the agent has learned about rocks and their categories.  The rock-to-feature table stores what the agent has learned about a given rock's features. When a fact is taught, a {\it note} is generated (e.g., ``Schist is a Metamorphic rock", ``Conglomerate has sand or pebbles") and automatically linked to a rock-to-category mapping, or one or more rock-to-feature mappings.  Additionally, if users provide an explanation in  \textit{Explain}, the explanation is added to the note as well (e.g., ``Granite has large crystals because the cooling process is slow''). When two rocks from the same category are compared in \textit{Compare}, the note associates these rocks to the same feature (e.g., ``Schist has layers and Gneiss has layers''). If the rocks are from different categories, then the note associates each rock to a different feature (e.g., ``Quartzite could be white because ... while Sandstone has layers'').  Finally, fun facts (from \textit{Fun Fact}) are stored along with the reasons that the users provided the agent as to why they found it interesting. An example is ``There is Gneiss in Canada that date back 4 billion years ago! (Reason: It is fascinating to know that rocks more than 4 billion years old can be found in this country)''.

\begin{table}[ht]
\caption{\textit{Correct} conversation flow in the second version of the Curiosity Notebook}
\label{tab:CorrectConversation}
\vspace{0.25cm}
{\small
\begin{tabular}{l p{11cm}}
{\bf Gamma:} & I got something wrong? \\ 
{\bf Gamma:} & I know 'gneiss', 'shale', 'slate' \\ 
{\bf Gamma:} & Ok, tell me which rock this is about? \\
{\bf User:} & Gneiss \\
{\bf Gamma:} & Oh, gneiss \\ 
{\bf Gamma:} & Select which notebook entry you want to correct! \\ 
{\bf User:} & I think that 'Gneiss is an igneous rock' is wrong. \\ 
{\bf Gamma:} & So, what kind of a rock do you think gneiss is then? \\
{\bf User:} & Metamorphic \\ 
{\bf Gamma:} & Thanks for clearing this up! \\
{\bf Gamma:} & Thanks for that information. Now you can select another button to keep teaching me!
\end{tabular}}
\end{table}
\vspace{0.25cm}

Students can click on the {\it Correct} button to update a fact that was previously taught to the agent. A sample conversation is a shown in Table \ref{tab:CorrectConversation}---the agent will first ask what object the student wants to focus on, then present the student with a list of learned facts about that object (entries in the agent's notebook) to choose from, and finally, use questions to elicit a specific kind of correction. 

In the first version's design, teaching is considered done when all the knowledge bubbles were filled. With the notebook, on the other hand, there is no simplistic idea of knowledge mastery. The users are told to teach the agent for 40-minutes in the way that they like, and are free to interpret if and how much the notebook is indicative of their competence as a teacher.  Having the notebook also gives users more transparency into how the agent learns, what it has been taught, and what it has yet to learn. As a more explicit representation of the agent's knowledge (compared to the knowledge bubbles in the first version), the platform better supports  metacognitive skills like identifying gaps and mistakes in the agent (and the student tutor's own) understanding.  Finally, the notebook feature presents a future opportunity to simulate the behaviour of different types of agent learners, e.g., an attentive agent learner might take notes of everything, whereas an inattentive learner might only occasionally take notes.

\subsubsection{Agent's Learning Mechanism}

In the first version, the agent is simplistic---it does not understand students’ responses to questions, and always pretends to learn what students have taught. Initially, the agent is unable to answer any test questions correctly; its ability to answer test questions increases with the number of features it has learned (i.e., number of completed teaching conversations). The second version improves upon this: the agent answers test questions according to the set of features it has been taught about the tested entity. As such, the agent is not able to categorize entities that it has not been taught before (and will inform the student tutor as such). It will also categorize an entity incorrectly if it has been taught the wrong information. As discussed previously, the second version's agent knowledge is represented using the agent's notebook in the frontend, and using the entity-to-category and entity-to-feature mappings in the backend. This allows the agent to answer test questions according to its knowledge, and allows the student tutors to accurately gauge the agent's learning.

\subsubsection{Admin Interfaces}

\begin{figure}[t!]
  \centering
  \includegraphics[height=9cm]{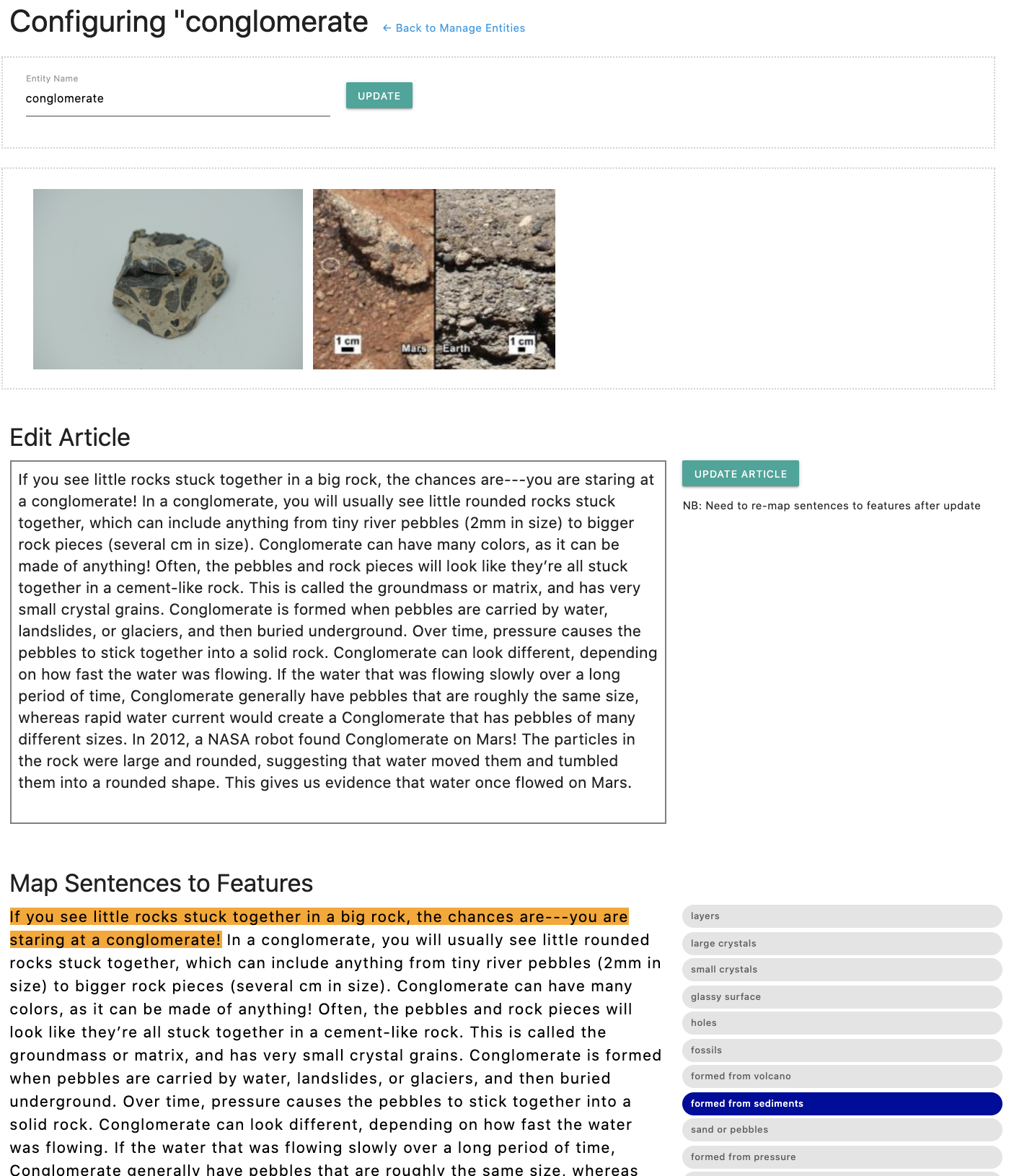}
  \caption{Admin interface for mapping sentences to relational facts, in the second version of the Curiosity Notebook. Right image courtesy NASA/JPL-Caltech/MSSS and PSI \cite{nasaRockPic}}
  \Description[Image of admin interface]{The picture shows the admin interface of Curiosity Notebook used for configuring articles. Researchers can edit the articles' images, text, and map each sentence in the article to one or more features.}
  \label{fig:admin_mapper}
\end{figure}

As we envision the Curiosity Notebook may be used by researchers to conduct studies and, eventually, by teachers to organize learning by teaching activities for their class, the platform provides a set of web-based administrative tools for adding/removing users, updating user information, assigning users to groups, as well as configuring classification tasks and materials (e.g., articles, images).  There is, for example, a semi-automatic way of mapping sentences to relational facts (e.g., Gabbbro is an igneous rock, Gabbro has holes), as shown in Figure \ref{fig:admin_mapper}.  For features, the system automatically maps sentences to features by scanning for a set of keywords or phrases (e.g., holes, bubble, bubbles, porous, cavities) that are synonymous to a feature (e.g., ``has holes''), and these mappings are subsequently manually verified, and corrected if needed, by the human user organizing the learning-by-teaching session. 
These admin features allow researchers and teachers to easily configure the learning material (CF\#3) and tutor groups (CF\#4), thus allowing teachers and experimenters to quickly adapt to unexpected changes during learning-by-teaching sessions, e.g., having to modify group assignments due to team dynamics problems or absent group members, or having to adapt materials on the fly.

As a research platform, the Curiosity Notebook provides researchers with the ability to configure experiments.  Functionalities include the ability to add/remove experiments, add/remove conditions to/from experiments, and assign users to a specific experiment and condition. The platform also provides
functionalities for researchers to configure the verbal behaviour of the agent, and associate different verbal behaviour with different experimental conditions (CF\#1), and deploy these agents, each with a distinctly different verbal behaviour, simultaneously to different participants.
Following, we describe findings from two deployments in different settings that were made possible due to Curiosity Notebook's support of the five CFs.


\section{Utility of Curiosity Notebook as a Research Platform for learning by teaching}\label{sec:utility}

In this section, we describe findings from the two deployments (detailed in Section \ref{sec:deployments}), as a way to demonstrate how the configurability of the Curiosity Notebook makes it a useful research platform for learning by teaching.  Our two deployments, which took place in contrasting learning environments---in one deployment, elementary school students taught a physical NAO robot in groups; in the second deployment, university students taught a text-based conversational agent individually---provide evidence for the versatility and utility of our platform.  The ability of our platform to adapt learning materials (CF\#3), group sizes (CF\#4) and agent embodiment (CF\#5) enabled us to quickly configure the system to deploy to vastly different learning settings.  In this section, we describe, in some details, the utility of certain configurable features and what they enabled us to observe during the two deployments.  For observations from the first deployment, we describe the results in aggregate as much of the observations were gathered informally from children. Due to the informal nature of the after-school club during which the sessions were run, children worked mostly with the same group of children, but were sometimes reassigned to a different group due to absences and personal preferences. For the second deployment which involved individual sessions, participants are denoted here as p1,...,pn, respectively. \\

\noindent {\bf The Impact of Agent Characteristics on Perception and Learning-by-Teaching Behaviour.}   Our findings suggest that different students perceived the same agent characteristics (e.g., programmed personality) in both positive or negative ways.  In the first deployment, some elementary school students attributed the behaviour of the agent as indicative of it being a good learner; students said that the agent is a good learner because of its attentiveness (e.g., ``because it pays very close attention'', ``because he/she ...sits in one spot and doesn't get distracted''), curiosity (e.g., ``because it’s curious'', ``because Delta asks questions, just like a human student''), and its ability and eagerness to learn (e.g., ``because he got everything right'', ``because he’s always ready to learn''). However, other students had more negative perceptions, mentioning the rationale of ``talking too much'' as the reason for the agent not being a good student.  Likewise, in the second exploratory study, most participants perceived Gamma as ``eager'' (p9, p11, p13, p15, p20, p22, p23, p31), ``enthusiastic'' (p13), ``positive'' (p15, p26, p40), ``pleasant'' (p16, p18), ``cheerful'' (p17, p24, p25), and ``excited'' (p16, p17, p36), and a good student because it takes notes. On the other hand, some participants found Gamma to be disingenuous, e.g., ``over enthusiastic'' (p33), ``a bit over the top'' (p35), ``fake'' (p3, p5), ``artificial'' (p6) and ``repetitive'' (p3). One participant (p21) suggested that a more natural version of Gamma would be one where ``she would stop paying attention if you don't use an entertainment button, and lose focus and maybe jots down the wrong answer.''

How students perceived the agent also critically affected how they perceived themselves as teachers.  One interesting observation from the first deployment was that what the agent said, as well as the degree to which the agent was reported to be a ``good'' student, seemed to be associated with students' perceptions of their own competence as teachers. That is, the majority of the students in the study saw themselves as good teachers and attributed their success at teaching to not only the learning progress of the robot (e.g., ``because my robot has learned a lot'', ``because we got all the bubbles for animals''), but also to the positive feedback they received from the robot (e.g., ``because the robot told me so'', ``because Delta always says good choice''). In contrast, university students in the second deployment, who had more access to information about what the robot has actually learned, seemed to rely more on accuracy information to judge their own competence as teachers. Participants who thought they taught well gave reasons including: 1) the notes written in the agent's notebook were all correct (p0, p6), 2) the agent answered all quiz questions correctly (p1, p7, p19, p20, p22, p24, p38, p40), and 3) they taught the agent all 12 rocks (p11, p13-16).  None of the participants, except for one (p20), mentioned the agent's conversational feedback helped them see themselves as a good teacher, even though both the first and second deployment agents were designed to have the same level of enthusiasm.  Given this observation, it may be beneficial for future versions to modify the verbal repertoire of the teachable agent to include specific feedback to the students about their teaching.

Configurability of agent characteristics (CF\#1) therefore can allow us to observe how the agent, as a tutee, can affect student's self-perception and subsequently their learning and teaching behaviour.\\    

\noindent {\bf Coordination of Group-Based Teaching}. The configurable feature CF\#4 allows us to create student tutor groups of varied sizes.   During the first deployment, students taught the agent in groups of 3.  This (i.e., having many students tutor the same agent tutee) is a somewhat atypical scenario in human-to-human peer teaching, but has been studied in human-to-machine teaching \cite{brezeal13, Hood2015}.  We observed that some students took the initiative to offer help to their teammates when it was not their turn to teach, while others were impatient at having to wait. Interestingly, the amount of attention that the robot gives to each student tutor seems to also affect students’ perceptions of their own teaching ability; one student said ``Student X teaches way better because the robot chooses X more.''  Together, these observations suggest the benefit of having a more idiosyncratic approach to managing group-based teaching that takes into account each student’s ability to work in a team and their unique need for attention from the agent.\\   

\noindent {\bf Quantification of Teaching Strategies}. Most importantly, our platform enabled us to observe, both qualitatively and quantitatively, {\it how} students go about teaching the agent (CF\#2).  

In the first deployment, we observed different groups of students demonstrating different teaching strategies, for example, some groups filled as many knowledge bubbles as possible before they tested the robot's knowledge; whereas other groups tested the robot often, e.g., after each filled bubble. Since the agent completely controls the conversation through a predefined sequence of questions, several students found this restriction as limiting their teaching process and wanted to be able to control how they teach more proactively. With the introduction of teaching buttons, the second deployment provides us with a much bigger opportunity to understand, quantitatively, the teaching behaviour of the students. 

For the second deployment, we performed a cluster analysis on the data from 40 participants (p10 was removed due to completing too few teaching conversations) to investigate whether participants' behaviour based on the amount of time they spent on different teaching activities reflected different teaching styles. There are 8 features in total---one for each of the 7 buttons, plus the button used to view the agent's notebook.   From these 8 features, 3 were removed upon further inspection; specifically, three buttons were used by participants less than three times on average throughout the entire teaching session: {\it Compare} ($M = 2.83$, $SD = 1.69$), {\it Correct} ($M = 1.1$, $SD = 1.19$) and {\it Tell Joke} ($M = 1.03$, $SD = 1.00$). In contrast, the 5 other buttons are used more frequently: \textit{Describe} ($M = 14.38$, $SD = 9.40$), \textit{Explain} ($M = 7.08$, $SD = 4.49$), \textit{Quiz} ($M = 6.03$, $SD = 3.07$) and \textit{Fun Fact} ($M = 4.55$, $SD = 2.99$). For each of these 5 teaching activities invoked by clicking button $X$, we create a feature to represent the relative attention that participants paid to that teaching activity, i.e., the number of times a participant clicked $X$ divided by the total number of button clicked over the entire session. Having 5 features is an appropriate amount according to Formann, who suggested no more than \textit{k} features for a sample size of $2^k$ when performing cluster analysis (i.e., $2^5 < 40$) \cite{formann1984latent}. The correlation between these 5 features were checked to ensure all features were distinctive during clustering \cite{hairMDA}. Hierarchical clustering was used as the clustering method; it was chosen given its use in past studies on tutors' behaviours on learning-by-teaching platforms \cite{kinnebrew2014analyzing}. The average silhouette width method \cite{rousseeuw1987silhouettes,10.1145/3322645.3322677} was used to analyze the quality of clustering. Its value ranges $[-1,1]$ and measures how close each point in their cluster is to other clusters, with a higher value indicating better quality clusters. Linear models were used to investigate the effects of other factors (demographics, perception of the agent, etc.) on button click rates. Stepwise selection was performed to select the variables that best explain the variability of each rate. The fit of the models were verified through visual inspection of the QQ-plots \cite{WILK1968}.

\begin{figure}[htbp!]
  \centering
  \includegraphics[width=0.3\linewidth]{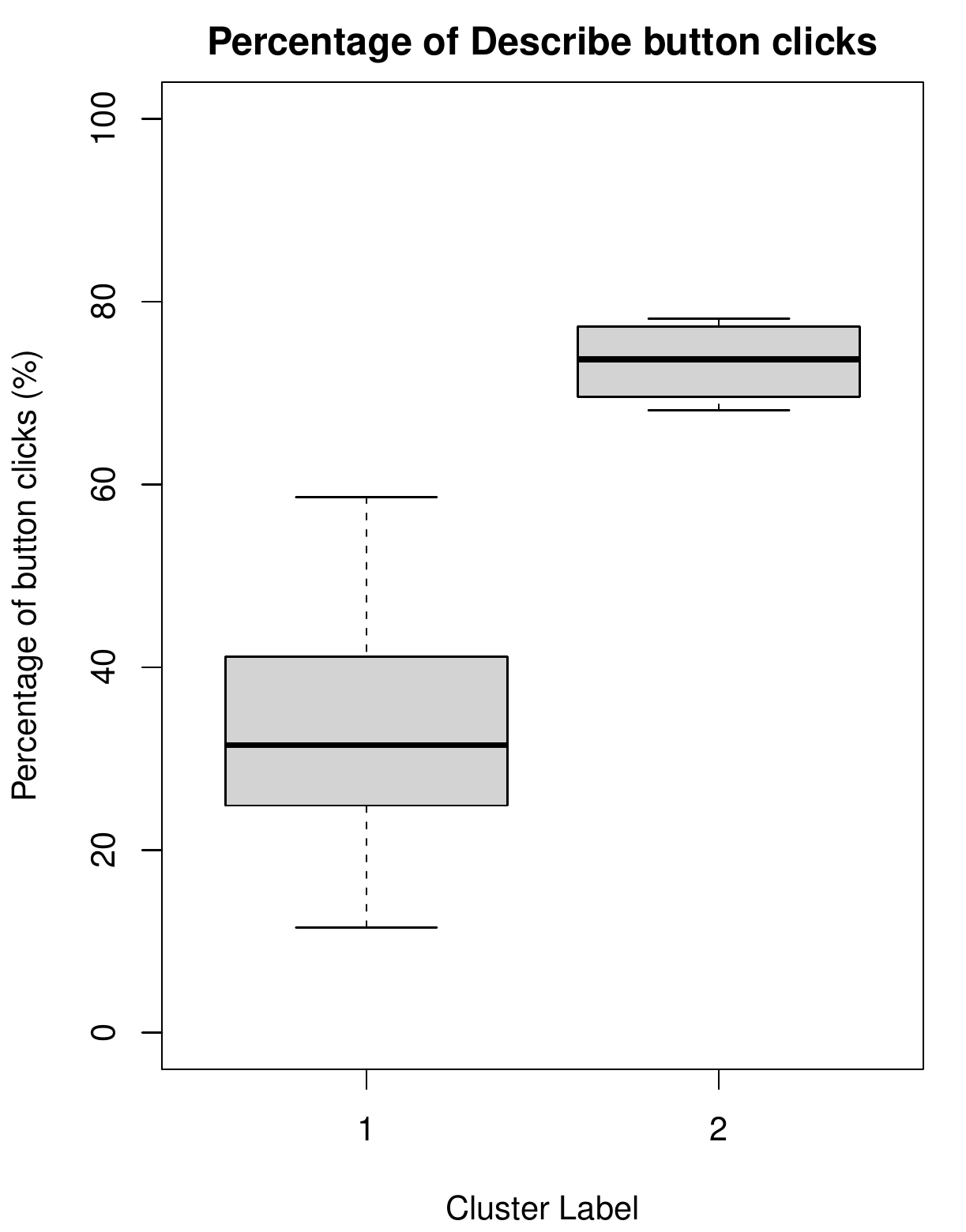}
  \includegraphics[width=0.3\linewidth]{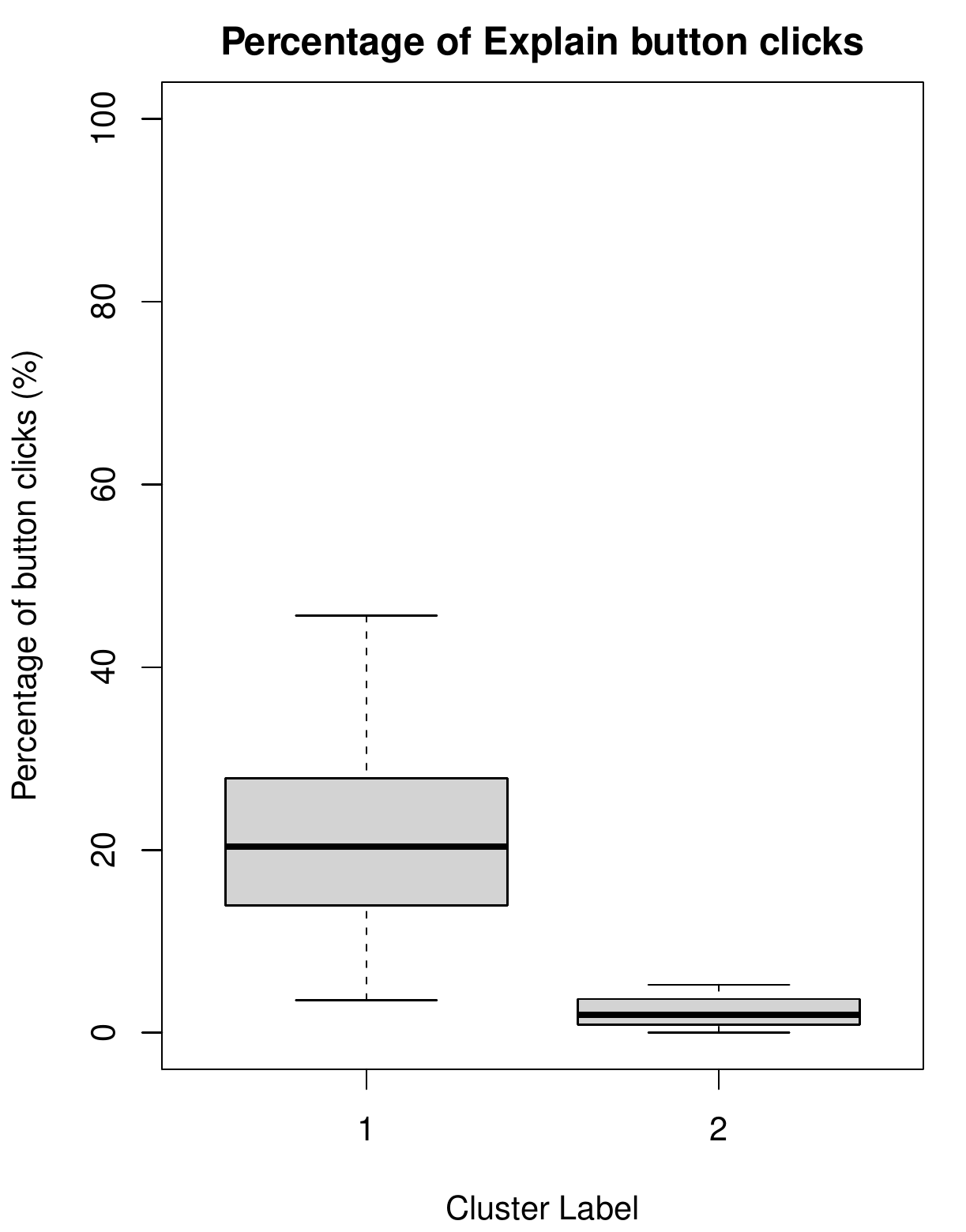}
  \includegraphics[width=0.3\linewidth]{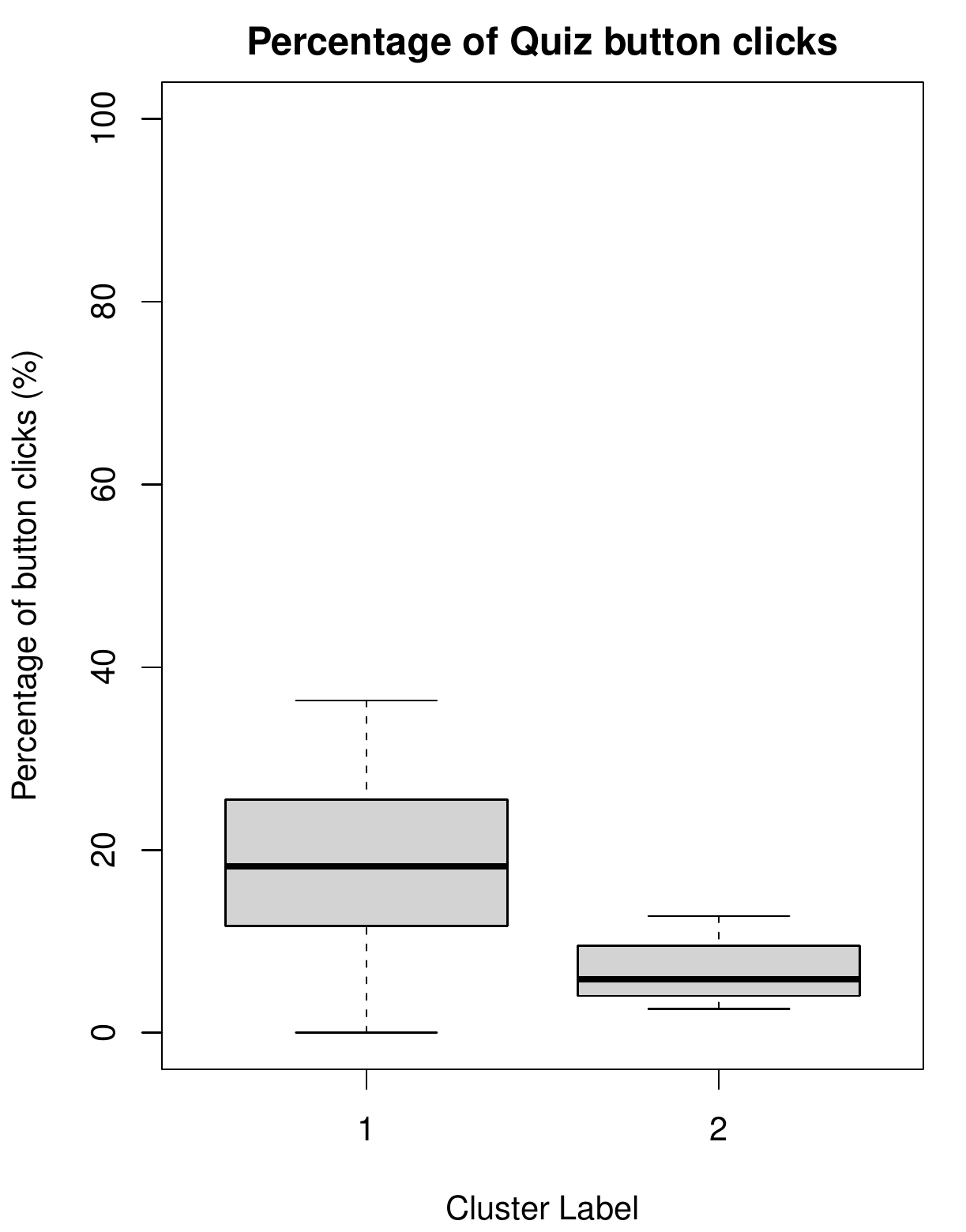}
  \includegraphics[width=0.3\linewidth]{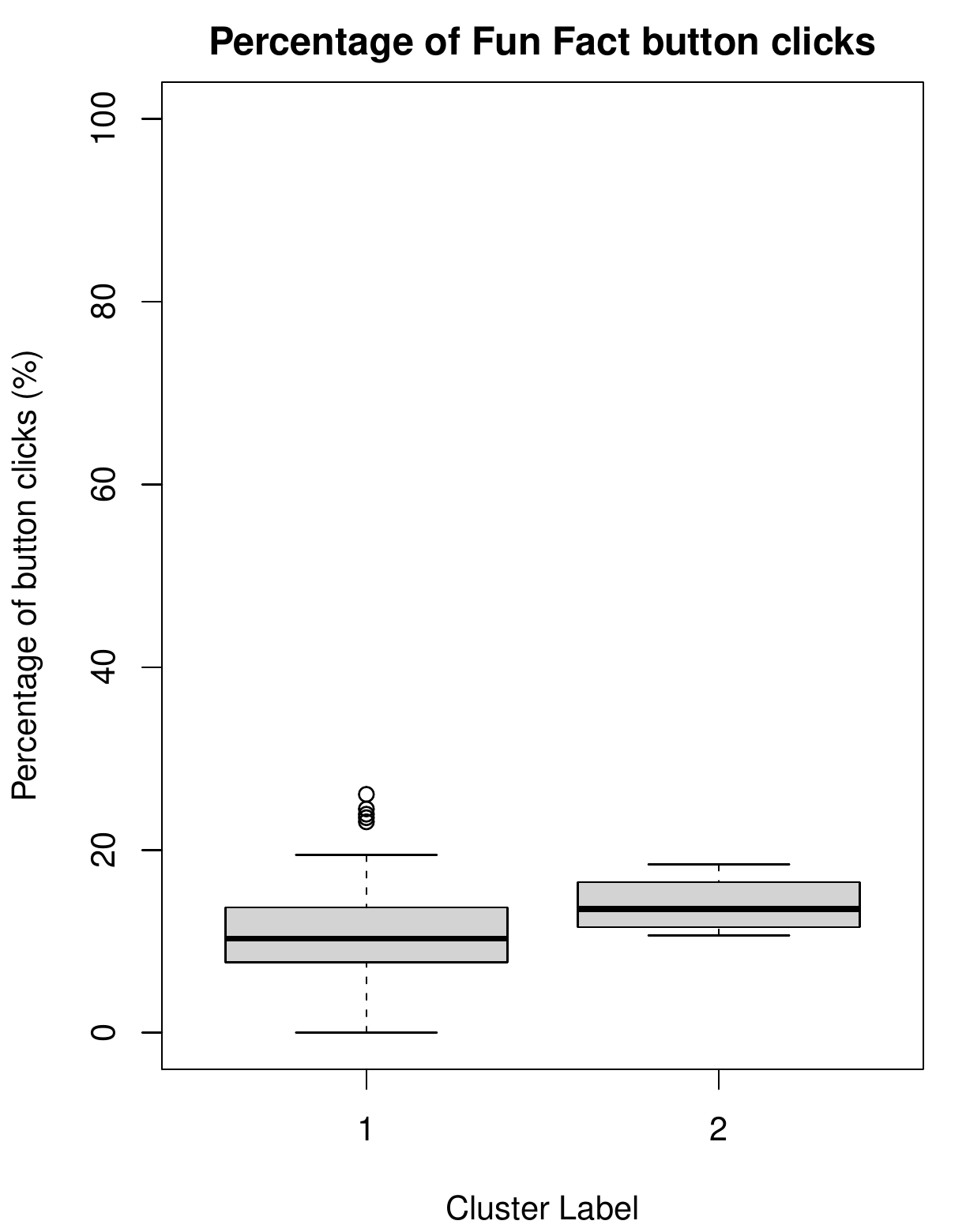}
  \includegraphics[width=0.3\linewidth]{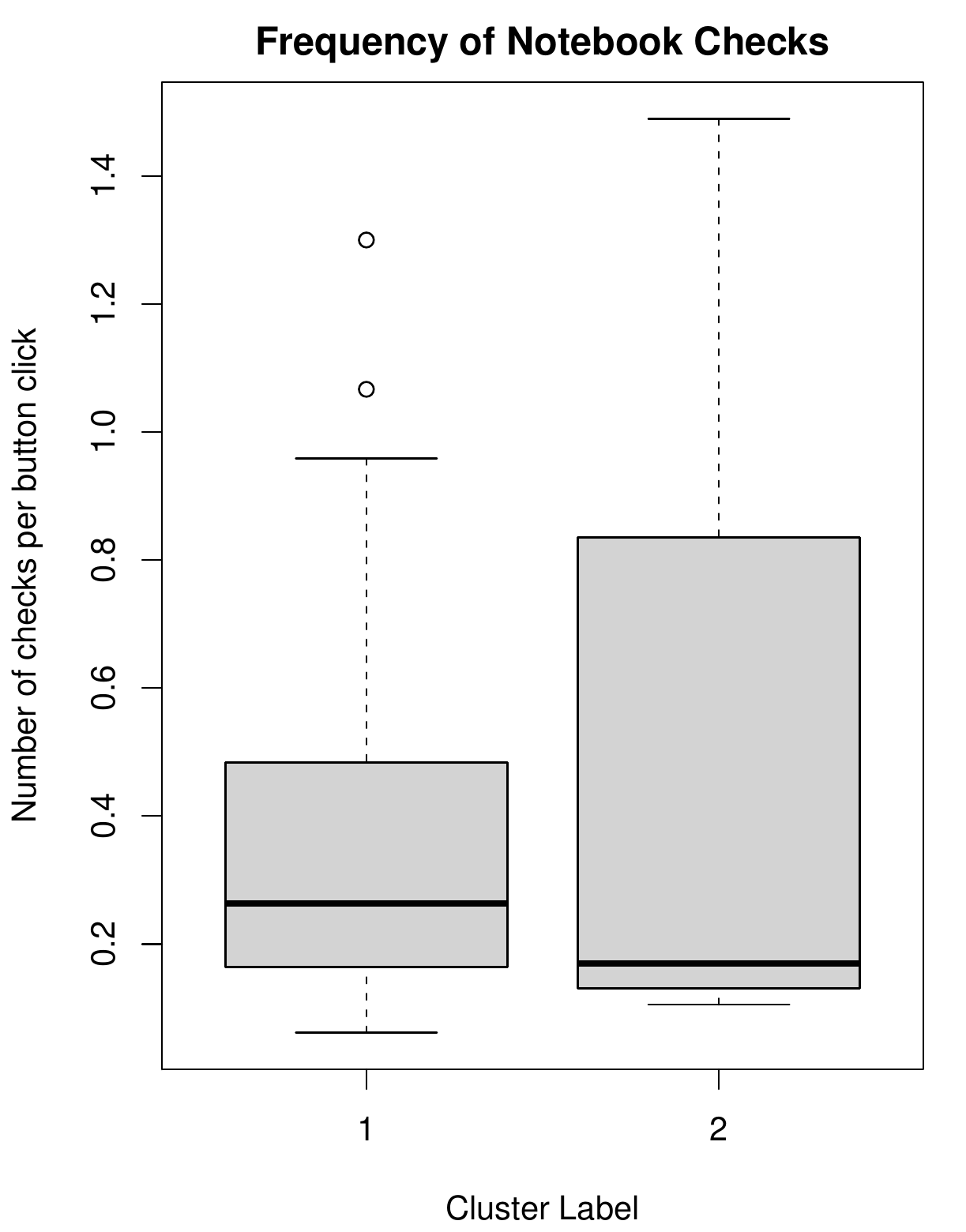}
  \caption{Comparison of clusters on features used for clustering (deployment 2).}
  \Description[Boxplots]{Four boxplots show the difference between clusters 1 and 2. The first boxplot shows cluster 1 having a higher percentage of teach type conversations. The second boxplot shows cluster 2 having a higher percentage of quiz conversations. The third boxplot shows cluster 2 having a higher rate of quiz conversations. The last boxplot shows cluster 2 having a higher average number of entertain type conversations between every describe conversation.}
  \label{fig:rq1_cluster}
\end{figure}

Two clusters were found ($N_1 = 36, N_2 = 4$), referred to as C1 and C2. These clusters achieve an average silhouette index of $0.36$ with all participants having positive index values, indicating that all participants fitted well within their assigned cluster. As shown in Figure \ref{fig:rq1_cluster}, participants in C1 have higher percentages of using {\it Explain} ($Mdn_1 = 20\%, Mdn_2 = 2\%$) and {\it Quiz} ($Mdn_1 = 18\%, Mdn_2 = 6\%$). They also check the agent's notebook more frequently ($Mdn_1 = 26\%, Mdn_2 = 17\%$). Participants in C2, on the other hand, have a much higher percentage use of {\it Describe} ($Mdn_1 = 31\%, Mdn_2 = 74\%$), and a slightly higher percentage use of {\it Fun Fact} ($Mdn_1 = 10\%, Mdn_2 = 14\%$).

Taking these results together, C1 participants can be interpreted as teachers who are more engaged than C2 participants.  On average, 87\% of button clicks by C2 are {\it Describe} and {\it Fun Fact} (which requires regurgitation, instead of synthesis, of facts), compared to only 42\% by C1. C2's teaching strategy is best described by p31, ``I mainly described things (rocks) to Gamma rather than explain''. On the flipside, C1 participants focus more on other aspects of teaching, like providing explanations via {\it Explain}, probing agent performance and gaining crucial feedback via {\it Quiz} and tracking the agent's learning by checking the notebook.  Fitting linear models on features beyond the final set of clustering features also revealed other significant differences.  For instance, C2 participants ($Mdn_2 = 68$) have, on average, significantly more notes than C1 participants ($Mdn_1 = 46$) with a large effect size $r = .45$ ($\beta = 36.28, t(37) = 5.86, p < .01$), as the {\it Describe} conversation is less time-consuming. This clustering is also in line with Felder and Silverman \cite{felder1988learning}, which discussed two different teaching styles---Concrete, which involves teaching in a repetitive manner, as exemplified by C2 participants; and Global, which is associated with a more diverse and creative teaching repertoire, as exemplify by C1 participants \cite{felder1988learning}. 

We were also interested in the characteristics of the two clusters, and thus examined participants' individual responses to the questionnaires. The AMS \textit{IM-to know} scores show that C1 participants are more intrinsically motivated to know ($Mdn_1 = 5$) than C2 participants ($Mdn_2 = 3.5$). A Mann-Whitney test shows significant difference, $U(N_1=36, N_2=4) = 131, p = .01$, and the r-value shows an effect size of $.42$, which is a large effect as suggested in \cite{fritz2012effect}. Also, the AMS \textit{EM-introjected regulation} scores show that C1 participants are more likely to internalize their teaching behaviour ($Mdn_1 = 4.33$) than C2 participants ($Mdn_2 = 1.67$); significance is found by a Mann-Whitney test, $U(N_1=36, N_2=4) = 124.5, p = .02$, alongside a large effect ($r = .38$). Cohen's \textit{d} also showed large effect sizes for both \textit{IM-to know} and \textit{EM-introjected regulation} score differences ($1.5$ and $1.6$ respectively) \cite{cohen1992power}.

These aspects have all been found to be beneficial not only to teaching, but also learning. C1 scores higher on the \textit{IM-to know} scale, which relates to the satisfaction of learning, exploring and understanding something new.  Through fitting Gaussian linear models, we also found that participants who had higher \textit{IM-to know} scores scored better in the post-study questionnaire for questions about rocks which do not have articles in the Curiosity Notebook $\beta = .42, z = .20, p = .03$. In other words, these participants scored significantly better on rocks that were not previously seen when teaching the agent; they were better at transferring classification rules learned during the teaching session to new contexts (i.e., new rocks). This suggests that participants with higher \textit{IM-to know} could end up learning better transfer skills than others due to their tendency to engage in teaching activities involving the integration of facts (as observed in C1).

In short, the two deployments demonstrate the Curiosity Notebook's configurable features (detailed in Section \ref{sec:essential_features}), and, crucially, the ability to generate important insights due to that flexibility. For instance, supporting configurable agent characteristics (CF\#1) provided information about the effects of the tutee agent's characteristics on students' behaviour and perceptions of themselves and the agent; supporting group-based teaching (CF\#4) confirms the need for more studies (that can be carried out using Curiosity Notebook) into understanding group dynamics during learning by teaching; providing a quantification of teaching behaviours (CF\#2) allowed for insights into less vs. more engaged participants' teaching behaviour and learning outcomes. Moreover, supporting configurable learning material (CF\#3) allowed for each deployment to be carried out for different age groups. Lastly, supporting flexible agent embodiment (CF\#5) allowed rapid pivoting from in-person studies, e.g., NAO robots in Deployment 1, to online studies, e.g., text-only agent in Deployment 2, vice versa.

\section{Envisioning Version 3: A Discussion}\label{sec:discussion}

In this section, we summarize Curiosity Notebook's design and changes from the first to second version in terms of the configurable features (CF). We also discuss potential improvements for the next version.

The configurability of the platform substantially improved from the first to second version.  In the first version, the agent was programmed to be an enthusiastic and quick learner, and its implementation did not allow researchers to easily configure agent characteristics (CF\#1). That is, they could not be swapped on-the-fly, nor could agents with different characteristics be deployed simultaneously (e.g., for between-subjects studies). Second, students were not given any autonomy over their teaching behaviour, such as the article taught and the way it is taught (CF\#2). Third, there was no easy way for researchers to modify articles or verify/correct the mapping between content (e.g., sentences) to concepts (e.g., categories, features) (CF\#3). Fourth, although the platform allowed for easy and quick allocation of students into groups of any size, more personalized methods for encouraging effective collaboration within groups should be considered (CF\#4). Lastly, following CF\#5, the platform enabled flexible agent embodiments. Compared to version 1, version 2 allowed for easy configuration of agent characteristics through the use of JSON configuration files (CF\#1). Second, it provided students with 7 buttons that initiated 7 distinct conversations (CF\#2), and the second deployment demonstrated the amount of flexibility allowed in teaching behaviour through the discovery of two distinct behaviours that affected learning. Third, new administrative interfaces allowed researchers to easily associate different agent characteristics with different experimental conditions (CF\#1), configure the material (CF\#3) and tutor grouping (CF\#4). 

These improvements allowed us to conduct a number of studies (described elsewhere in \cite{cehaHumour,ravarieffects}) where the agent's characteristics were the experimentally manipulated through quick configurations.  It allowed us to release the second deployment easily, even though a complete overhaul of the material was required. Moreover, it allowed us to collect detailed quantitative data on how students go about teaching. For CF\#2, a potential improvement is to design the platform to collect more detailed data on tutor-agent interaction, so that complex interactions, such as those mentioned by Roscoe and Chi, can be appropriately captured and analyzed \cite{Roscoe2007}. In terms of CF\#4, we envision the next version of the Curiosity Notebook to include configurability on how teams of students communicate with each other, and with the agent. For instance, a separate JSON file could be used to list preferred details of various aspects of group communication, such as how frequently should the agent explicitly encourage discussions between group members, how frequently should the agent refer to material taught by a group member while being taught by another, and what (if any) ways can group members interfere when a member is teaching. Further improvements can also allow for the integration of machine learning models that impact either agent characteristics (CF\#1) or group communication (CF\#4). In terms of CF\#5, the next version could provide more standardized application programming interfaces (API) to enable any output devices to act as an agent. This would enhance flexibility by further reducing the ease of connecting various software (e.g., text-to-speech software) or hardware (e.g., robots) to the platform.

One of the less satisfying aspects about the learning-by-teaching platform so far is our ability to demonstrate learning gains in the students themselves.  A potential reason is that the learning objectives of the agent is too inconspicuous---the {\it Quiz} action, which demonstrates how well the agent can classify objects, needs to be invoked explicitly by the student tutor.  Prior research has showed that participants learn better when observing their students use the knowledge they taught \cite{Okita2013, Okita2013_2}. In our studies, we asked participants to teach the agent, but we did not tell them {\it why} the agent needs to acquire that knowledge. The student's sense that the knowledge taught to the agent matters could become a source of extrinsic motivation for the user, further enhancing their learning. Thus, having the teachable agent demonstrate its skills openly to students while it is being taught can be both informational (i.e., shows where the misunderstandings lie) and motivational.  In version 3 of the Curiosity Notebook, we envision adding a ``demonstration panel'', where the agent is performing a task live (e.g., sorting objects into different categories, writing code, describing the steps to put together a recipe) while soliciting feedback and help from the students.  Having the agent show its mistakes in an obvious fashion can also prompt for specific forms of teaching, without needing explicit teaching buttons; for example, students may be inclined to teach certain rules to the agent in order to fix its mistakes or to help it reach its goal. 

There are a few limitations in our design approach. The first deployment was over 4 weeks, thus allowing us to observe learning-by-teaching behaviours over time.  However, the sample size was considerably small, which limited the potential for more complex analysis on participants' teaching behaviours.  Additionally, since the design of the two deployments were not consistent, no direct comparison could be made between the studies.  Nonetheless, the two deployments and multiple design iterations allowed us to understand the intricacies of the learning-by-teaching process and what is required for a research platform for learning by teaching to be maximally useful.


\section{Conclusions}
In this paper, we introduce the Curiosity Notebook, an interface that provides various teaching tools for participants to interact with a teachable agent. We iteratively designed the platform based on observations from two deployments---a field study with 12 elementary school children and an online study with 41 university students. We showed that by providing ways to configure features that were identified by prior research to be important, two learning-by-teaching studies deployed under very different contexts could be conducted on the same platform without much overhead. The Curiosity Notebook's value as a research utility was demonstrated by the important insights gained from the studies. 

Moving forward, there are plans to allow the research community to fully exploit Curiosity Notebook's utility as a research platform through open sourcing. This greatly reduces the overhead of conducting such research by removing the need to build platforms from the ground up. Moreover, having a platform that many can use for their research will allow for more direct comparisons in findings between various learning-by-teaching studies. This will make it much easier for findings to be generalized or aggregated to a level suitable to be applied directly in the real world by those in the education sector.

\section{Acknowledgments}
We thank all participants for their contributions, and acknowledge the funding from the NSERC Discovery Grant RGPIN-2015-0454 and the University of Waterloo Interdisciplinary Trailblazer Fund for making this work possible.

\bibliographystyle{ACM-Reference-Format}
\bibliography{main}

\received{October 2020}
\received[revised]{April 2021}
\received[accepted]{July 2021} 

\end{document}